\documentclass[aps,twocolumn,superscriptaddress]{revtex4-1}

\usepackage[utf8]{inputenc}
\usepackage[english]{babel}
\usepackage{amsmath}
\usepackage{amsfonts}
\usepackage{amssymb}
\usepackage{makeidx}
\usepackage{graphicx}
\usepackage{lmodern}
\usepackage{kpfonts}
\usepackage[left=1.5cm,right=1.5cm,top=2cm,bottom=2cm]{geometry}

\usepackage{xcolor}
\usepackage{hyperref}
\hypersetup{linkbordercolor=cyan}

\usepackage{amsmath}
\newcommand*\diff{\mathop{}\!\mathrm{d}}

\usepackage{verbatim}
\usepackage{caption}
\usepackage{subcaption}
\usepackage{tabularx}
\usepackage{floatrow}

\usepackage{sidecap}

\usepackage[margin=0.2cm]{subcaption}
\usepackage[toc,page]{appendix}
\usepackage{xcolor} 
\usepackage{graphicx}
\usepackage[export]{adjustbox}
\graphicspath{
{C:/Users/Utilisateur/Documents/Uni/these/latex/images/}
{C:/Users/Utilisateur/Dropbox/images/}
{C:/Users/Utilisateur/Dropbox/images/Langer/isotropic_annihilation/}
{C:/Users/Utilisateur/Dropbox/images/Langer/boundary_annihilation/}
{C:/Users/Utilisateur/Dropbox/images/Langer/2skyrmions_annihilation/}
{C:/Users/louise/Dropbox/images/}
{C:/Users/louise/Dropbox/images/screencaps/}
{C:/Users/louise/Dropbox/images/Langer/isotropic_annihilation/}
{C:/Users/louise/Dropbox/images/Langer/boundary_annihilation/}
{C:/Users/louise/Dropbox/images/Langer/2skyrmions_annihilation/}
{C:/Users/louise/Dropbox/images/Langer/}
{C:/Users/louise/Dropbox/images/Langer/non_mag_defect/}
{C:/Users/louise/Dropbox/images/Langer/blochsk_collapse/}
{C:/Users/louise/Dropbox/images/Langer/antisk_collapse/}
{C:/Users/louise/Dropbox/images/frustrated/sk_anni/}
{C:/Users/louise/Dropbox/images/frustrated/antisk_anni/}
{C:/Users/louise/Dropbox/images/frustrated/zoology/}
{C:/Users/louise/Dropbox/images/frustrated/Q2Q1/}
{C:/Users/louise/Dropbox/images/frustrated/Q2_Q1_Q1/}
{/home/desplat/Dropbox/images/Langer/boundary_annihilation/}
{/home/desplat/Dropbox/images/Langer/isotropic_annihilation/}
{/home/desplat/Dropbox/images/Langer/2skyrmions_annihilation/}
{/home/desplat/Dropbox/writing/latex/images/}
{/home/desplat/Dropbox/images/Langer/}
{/home/louise/Dropbox/writing/latex/images/}
{/home/louise/Dropbox/images/}
{/home/louise/Dropbox/images/screencaps/}
{/home/louise/Dropbox/images/Langer/isotropic_annihilation/}
{/home/louise/Dropbox/images/Langer/boundary_annihilation/}
{/home/louise/Dropbox/images/Langer/2skyrmions_annihilation/}
{/home/louise/Dropbox/images/Langer/antisk_collapse/}
{/home/louise/Dropbox/images/Langer/blochsk_collapse/}
{/home/louise/Dropbox/images/Langer/non_mag_defect/}
{/home/louise/Dropbox/images/Langer/}
{/home/louise/Dropbox/images/Langer/curved_boundary/}
{/home/louise/Dropbox/images/frustrated/sk_anni/}
{/home/louise/Dropbox/images/frustrated/antisk_anni/}
{/home/louise/Dropbox/images/frustrated/zoology/}
{/home/louise/Dropbox/images/frustrated/Q2Q1/}
{/home/louise/Dropbox/images/frustrated/Q2_Q1_Q1/}
{/home/louise/Dropbox/images/frustrated/}
{/home/louise/Dropbox/images/frustrated/Qm2/}
}

\usepackage{multirow}

\begin{document}
\title{Paths to annihilation of first- and second-order (anti)skyrmions \\ via (anti)meron nucleation on the frustrated square lattice}%


\author{L. Desplat}%
\email{l.desplat.1@research.gla.ac.uk}
\affiliation{Centre for Nanoscience and Nanotechnology, CNRS, Université Paris-Sud, Université Paris-Saclay, 91120 Palaiseau, France}
\affiliation{SUPA School of Physics and Astronomy, University of Glasgow, G12 8QQ Glasgow, United Kingdom}

\author{J.-V. Kim}
\affiliation{Centre for Nanoscience and Nanotechnology, CNRS, Université Paris-Sud, Université Paris-Saclay, 91120 Palaiseau, France}

\author{ R. L. Stamps}
\affiliation{SUPA School of Physics and Astronomy, University of Glasgow, G12 8QQ Glasgow, United Kingdom}
\affiliation{Department of Physics and Astronomy, University of Manitoba, Winnipeg, Manitoba, R3T 2N2 Canada}

\date{\today}

\begin{abstract}
We study annihilation mechanisms of small first- and second-order skyrmions and antiskyrmions on the frustrated $J_1-J_2-J_3$ square lattice with broken inversion symmetry (DMI). We find that annihilation happens via the injection of the opposite topological charge in the form of meron or antimeron nucleation. Overall, the exchange frustration generates a complex energy landscape with not only many (meta)stable and unstable local energy solutions, but also many possible paths connecting them. Whenever possible, we compute the activation energy and attempt frequency for the annihilation of isolated topological defects. In particular, we compare the average lifetime of the antiskyrmion calculated with transition state theory with direct Langevin simulations, where an excellent agreement is obtained.
\end{abstract}

\maketitle


\section{Introduction}
Magnetic skyrmions are solitonic, particlelike, topologically nontrivial magnetic textures. The existence of (meta)stable skyrmionic solutions in a system requires the introduction of a characteristic length-scale via competing interactions. This is typically achieved in noncentrosymmetric magnets with the Dzyaloshinskii-Moriya interaction (DMI) \cite{dzyaloshinskii,moriya}. That particular type of solutions, commonly referred to as chiral skyrmions, were theoretically investigated in the 1990s \cite{bogdanov1989thermodynamically, bogdanov1994thermodynamically}. In magnets with inversion symmetry, skyrmions can be stabilized via dipolar interactions (skyrmion bubbles) \cite{chikazumi2009physics,malozemoff1979magnetic,yu2012magnetic,finazzi2013laser,nagao2013direct,koshibae2016theory}, as well as frustrated exchange couplings  \cite{okubo2012multiple,nagaosa2013topological,lin2016ginzburg,rozsa2017formation,zhang2017skyrmion}. Systems with frustrated exchange are particularly interesting, as they allow many different kinds of topological defects to coexist. The frustrated $J_1-J_2$ system on the hexagonal lattice was extensively investigated in Ref.  \cite{leonov2015multiply} and, in general, the frustrated hexagonal lattice is the most commonly studied \cite{okubo2012multiple,dupe2014tailoring,von2017enhanced,rozsa2017formation,ritzmann2018trochoidal}. 
Skyrmions decouple from the lattice in the long wavelength limit and therefore, in that case, the choice of the lattice does not matter \cite{lin2016ginzburg}. 
Collapse mechanisms of skyrmions and antiskyrmions in chiral thin films with frustrated exchange were previously investigated \cite{bessarab2018annihilation,von2017enhanced}, and appear similar to the isotropic-type of collapse calculated in non-frustrated chiral systems \cite{gneb,lobanov2016mechanism,uzdin2017effect,cortes2017thermal,stosic2017paths,desplat2018thermal}. A third mechanism involving the injection of singularity with opposite topological charge was reported in \cite{cortes2017thermal}, but the path appears to go over a higher order saddle point. Additionally, this mechanism seems to emerge for larger skyrmion sizes, yet it is perhaps the closest to the type of mechanisms we report in this work.

In this article, we look at paths to annihilation of first- and second-order skyrmions and antiskyrmions on the frustrated  $J_1-J_2-J_3$ square lattice previously studied by Lin \textit{et al.} \cite{lin2016ginzburg}. We relax skyrmionic solutions spanning over only a few sites in diameter that do not exhibit translational invariance on the lattice, and instead experience pinning at particular lattice points. In the present system, the solutions are too small to decouple from the underlying lattice. In this context, we obtain new types of collapse mechanisms which have, to date, not been reported. Interfacial DMI is added to break the invariance with respect to the rotation of helicity (i.e. the ``spin" degree of freedom \cite{lin2016ginzburg}), as well as the degeneracy between skyrmions and antiskyrmions.

The paper is organized as follows. In Sec. \ref{sec:model}, we present the model Hamiltonian and some useful definitions, as well as a brief zoology of some of the different topological defects that can be stabilized in the system. In Sec. \ref{sec:results}, we firstly give regions of existence of metastable first- and second-order skyrmions and antiskyrmions in parameter space. We then present different mechanisms by which they annihilate, and we confirm them by means of direct Langevin simulations. We subsequently compute attempt frequencies associated with the different mechanisms via transition state theory calculations. Finally, the results are discussed in Sec. \ref{sec:discussion}.

 
\section{Model}\label{sec:model}
We simulate a bidimensional square lattice of $N$ magnetic spins $\{\mathbf{m}_i\}$ with a constant magnitude that we set to unity. We use the classical Heisenberg model Hamiltonian,

\begin{equation}\label{eq:frustr_hamiltonian}
	E = -\sum_{\langle ij\rangle} J_{ij} \mathbf{m}_i \cdot \mathbf{m}_j - \sum_{\langle ij\rangle} \mathbf{D}_{ij} \cdot \big( \mathbf{m}_i \times \mathbf{m}_j \big)  -  K \sum_i m_{z,i}^2 -B_z \sum_i m_{z,i},
	\end{equation} 	
where $J_{ij}$ is the strength of the isotropic exchange coupling extended to third nearest neighbors ($J_1-J_2-J_3$),  $\mathbf{D}_{ij}$ is the Dzyaloshinskii vector restricted to first nearest neighbors, $K$ is the perpendicular, uniaxial anisotropy constant, and $B_z$ is the perpendicular applied magnetic field. The DMI is interfacial and favors Néel skyrmions, in which case the Dzyaloshinskii vector is  defined as $\mathbf{D_ {ij}} = D \mathbf{r}_{ij} \times \mathbf{e}_z$, where $\mathbf{r}_{ij} $ is the in-plane direction between sites $i$ and $j$ \cite{thiaville2012dynamics}. We introduce the reduced parameters: $j_{\alpha} = J_{\alpha} / J_1, (\alpha=1,2,3)$ ; $d = \lvert \mathbf{D}_{ij} \lvert / J_1 $; $k = K / J_1$ ; $b = B_z / J_1$.  We set $k=0.1$ for the rest of this study. The isotropic exchange coupling parameters allow a spin-spiral ground state to be realized in the absence of other interactions, and are the following \cite{lin2016ginzburg}: $j_2 = -0.35$ ; $j_3 = -0.15$. Additional parameter values are given in Appendix \ref{app:parameters}. We define the topological charge \cite{belavin1975aa},
\begin{equation}\label{eq:topo_charge}
N_s =  \int \diff\mathbf{r}^2 \rho_s = -n,
\end{equation}
in which $\rho_s$ is the toplogical charge density defined as,
\begin{equation}\label{eq:topo_charge_dens}
\rho_s = \frac{1}{4\pi} \mathbf{m} \cdot (\partial_x \mathbf{m} \times \partial_y \mathbf{m} ).
\end{equation}

\begin{figure}[hbtp]
\centering	
		\begin{subfigure}[t]{.25\textwidth}		
	\adjincludegraphics[width=1\linewidth,trim={3cm 3cm 3cm 3cm},clip]{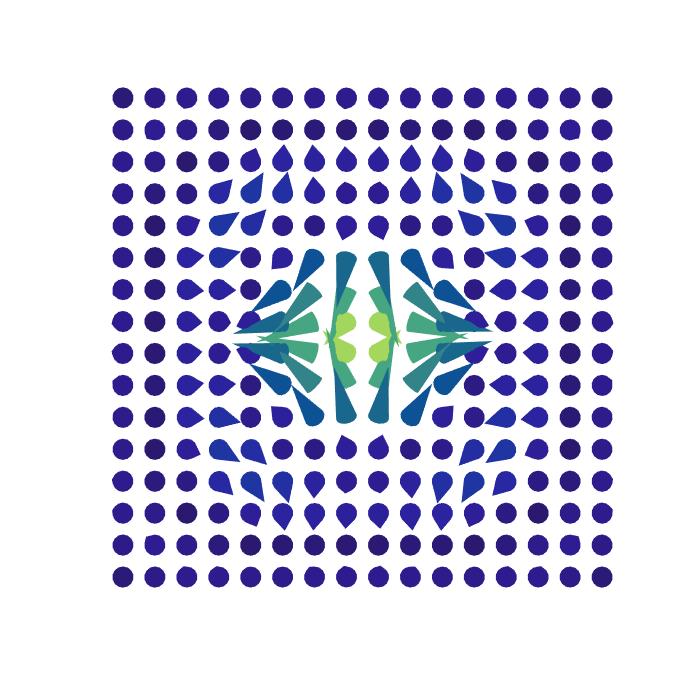}	
		\caption{}
		\label{fig:antisk}
	\end{subfigure}
		\begin{subfigure}[t]{.25\textwidth}		
			\adjincludegraphics[width=1\linewidth,trim={3cm 3cm 3cm 3cm},clip]{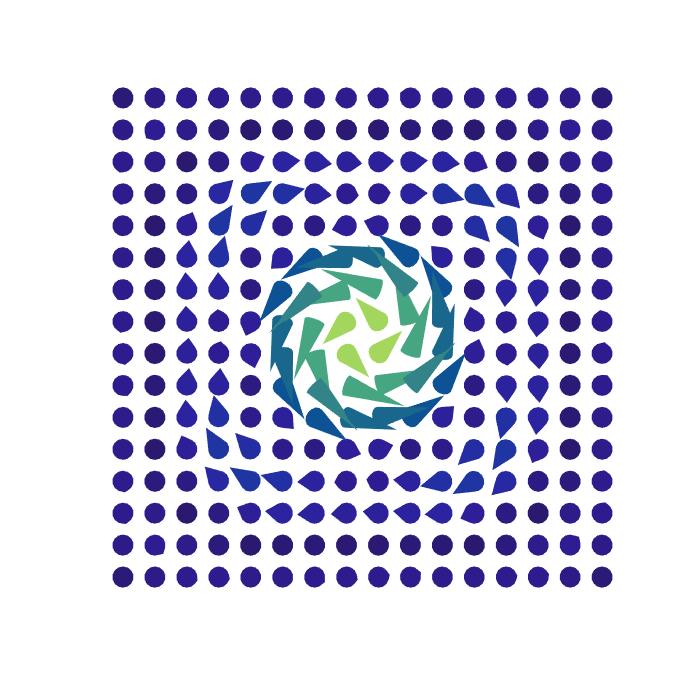}
		\caption{}
		\label{fig:blochsk}
	\end{subfigure}		
	\begin{subfigure}[t]{.25\textwidth}		
			\adjincludegraphics[width=1\linewidth,trim={3cm 3cm 3cm 3cm},clip]{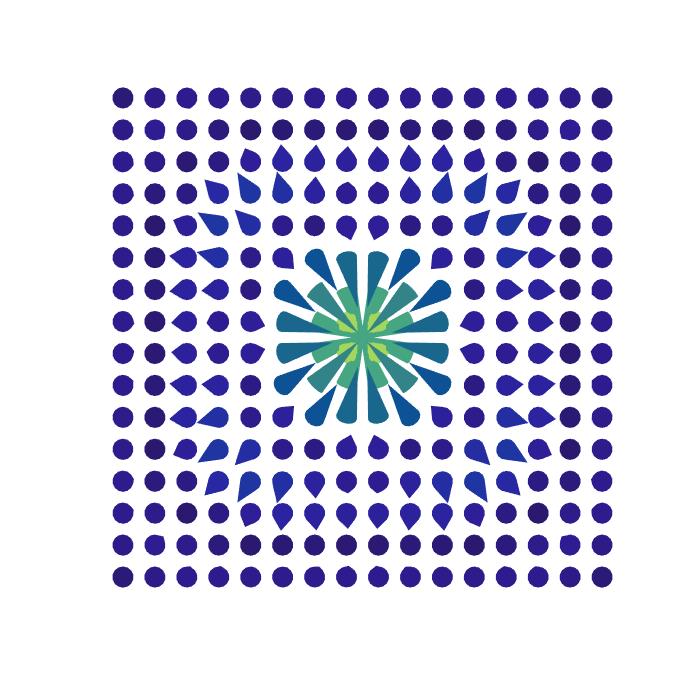}
		\caption{}
		\label{fig:neelsk}
	\end{subfigure}
	
		\begin{subfigure}[t]{.28\textwidth}		
			\adjincludegraphics[width=1\linewidth,trim={3cm 3cm 3cm 3cm},clip]{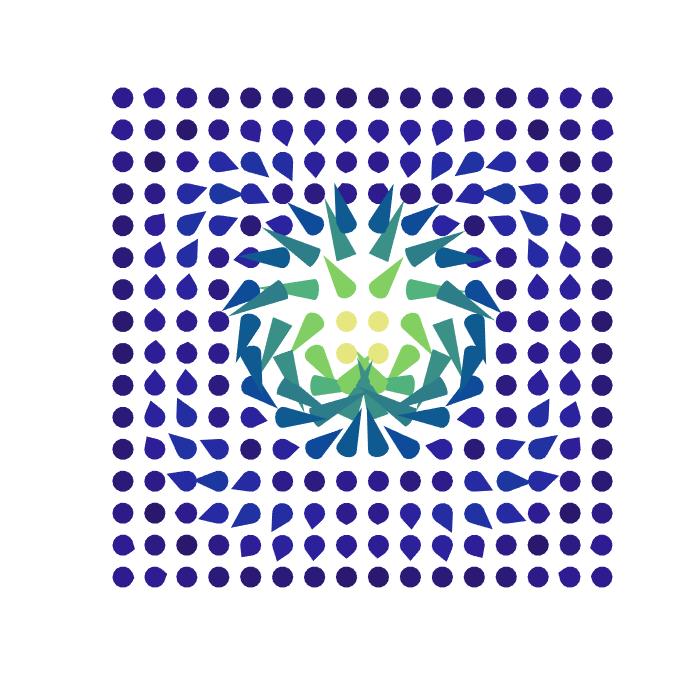}
		\caption{}
		\label{fig:Q2sk}
	\end{subfigure}		
			\begin{subfigure}[t]{.28\textwidth}		
			\adjincludegraphics[width=1\linewidth,trim={3cm 3cm 3cm 3cm},clip]{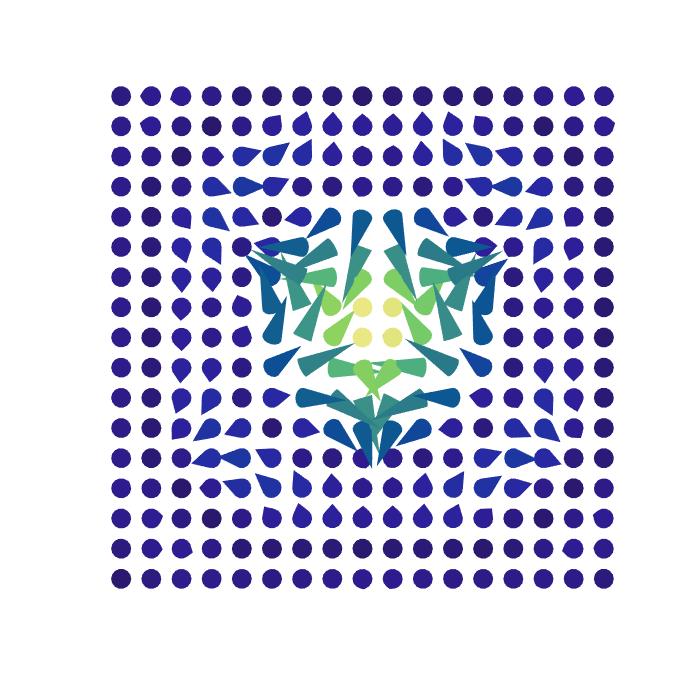}
		\caption{}
		\label{fig:Q2ask}
	\end{subfigure}\hfill
	
	\begin{subfigure}[t]{.2\textwidth}
				\includegraphics[width=1\textwidth]{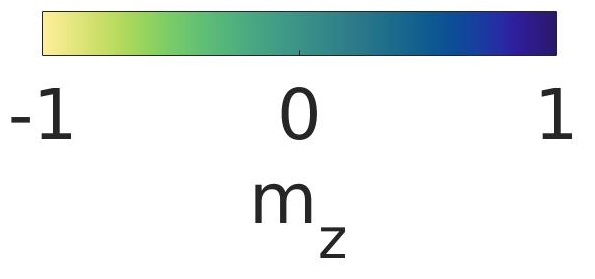}
	\end{subfigure}	
\caption{Different isolated topological defects stabilized in a frustrated magnet with inversion symmetry $(d=0)$ and reduced parameters $(b,k)=(0.1,0.1)$. (a) Antiskyrmion $(n = -1)$. (b) Skyrmion $(n = 1)$ (Bloch). (c) Skyrmion $(n = 1)$ (Néel). (d) Second-order skyrmion $(n = 2)$. (e)  Second-order antiskyrmion $(n = -2)$.  }
\label{fig:zoology}
\end{figure}

The quantity $n$ in Eq. (\ref{eq:topo_charge}) is the winding number, or vorticity, and
counts the number of times the spin configuration wraps the unit sphere $S^2$. 
 Solutions with a positive winding number are called skyrmions, and solutions with a negative winding number are called antiskyrmions.Contrary to chiral skyrmions stabilized by DMI where the type of the DMI favors one particular class of solutions, exotic spin textures with various winding numbers and helicities can coexist in frustrated magnets \cite{leonov2015multiply,lin2016ginzburg, rozsa2017formation,zhang2017skyrmion} [Fig. \ref{fig:zoology}].
Nonetheless, the possibility to stabilize structures with an arbitrary integer topological charge in chiral magnets (``skyrmionic sacks")  has recently been explored \cite{rybakov2019chiral}.
 In the absence of inversion symmetry breaking, skyrmions [Figs. \ref{fig:blochsk}, \ref{fig:neelsk}] and antiskyrmions [Fig. \ref{fig:antisk}] with opposite winding numbers are degenerate in energy. Additionally, magnets with inversion symmetry possess a spin degree of freedom (where ``spin" refers to the spin of the skyrmion as a particle), associated with a rotation of helicity.
Bloch-type skyrmions [Fig. \ref{fig:blochsk}], Néel-type skyrmions [Fig. \ref{fig:neelsk}], and all intermeditate helicity states are therefore degenerate in energy \cite{lin2016ginzburg}.

\begin{figure}[hbt!]
\centering
\begin{subfigure}[t]{.7\textwidth}		
		\adjincludegraphics[width=1\textwidth,trim={{.15\width} 0 0 {.0\height}},clip]{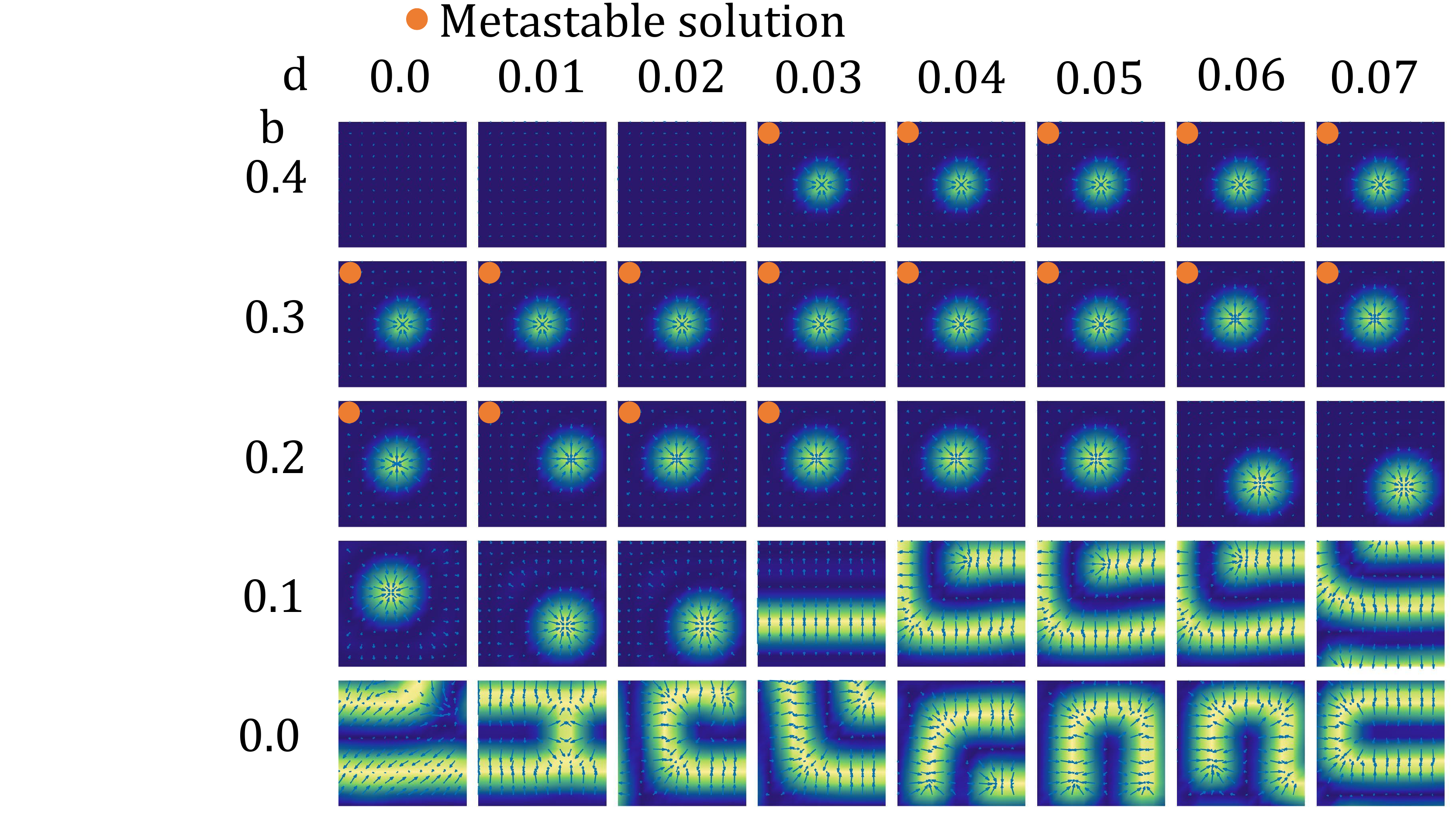}
		\caption{}
		\label{fig:bd_sk}
	\end{subfigure}
	
	\begin{subfigure}[t]{.7\textwidth}		
		\adjincludegraphics[width=1\textwidth,trim={{.15\width} 0 0 {.05\height}},clip]{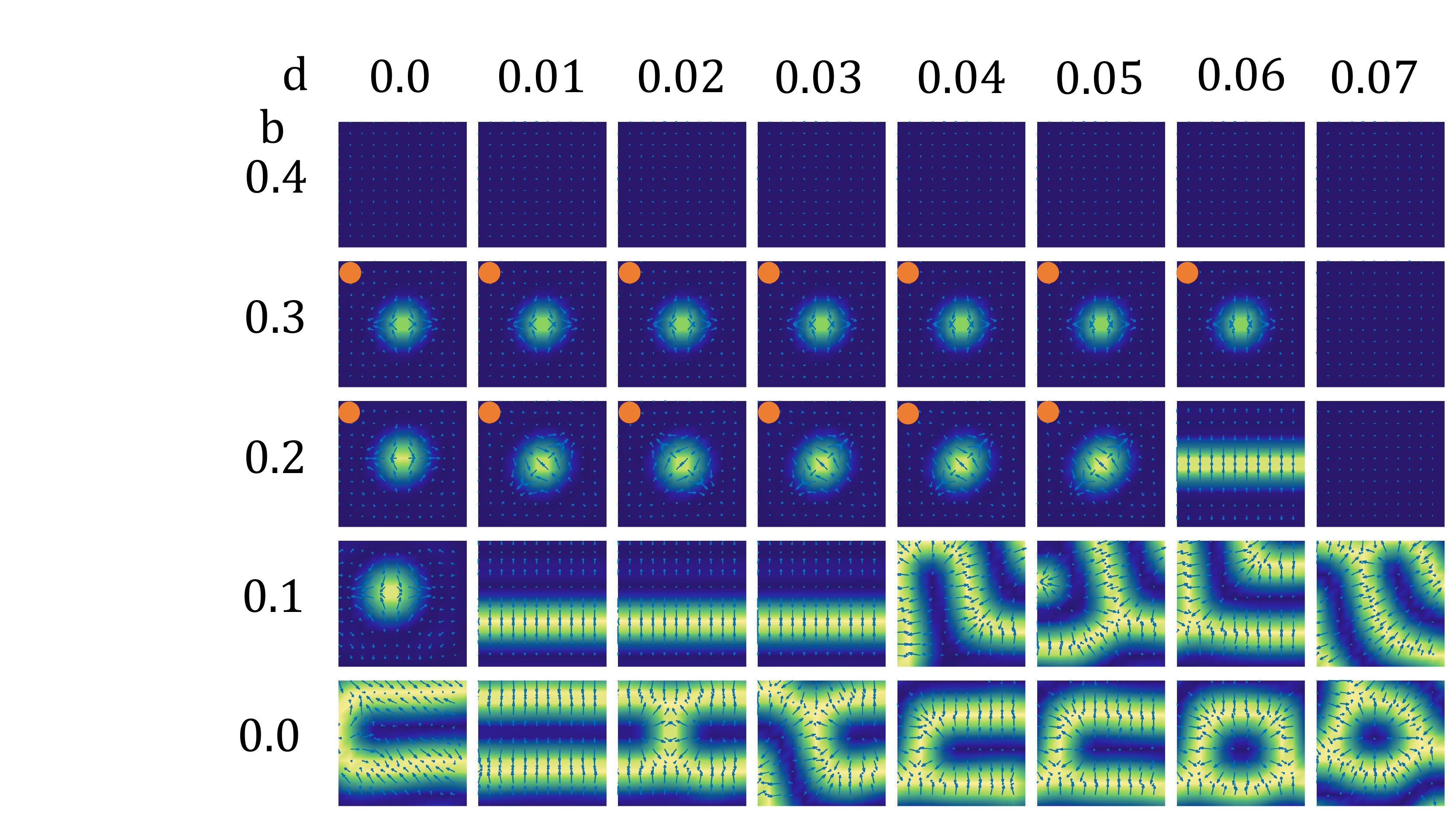}
		\caption{}
		\label{fig:bd_antisk}
	\end{subfigure}
	
	\begin{subfigure}[t]{.45\textwidth}	
			\adjincludegraphics[width=1\textwidth,trim={{.3\width} 0 0 {.05\height}},clip]{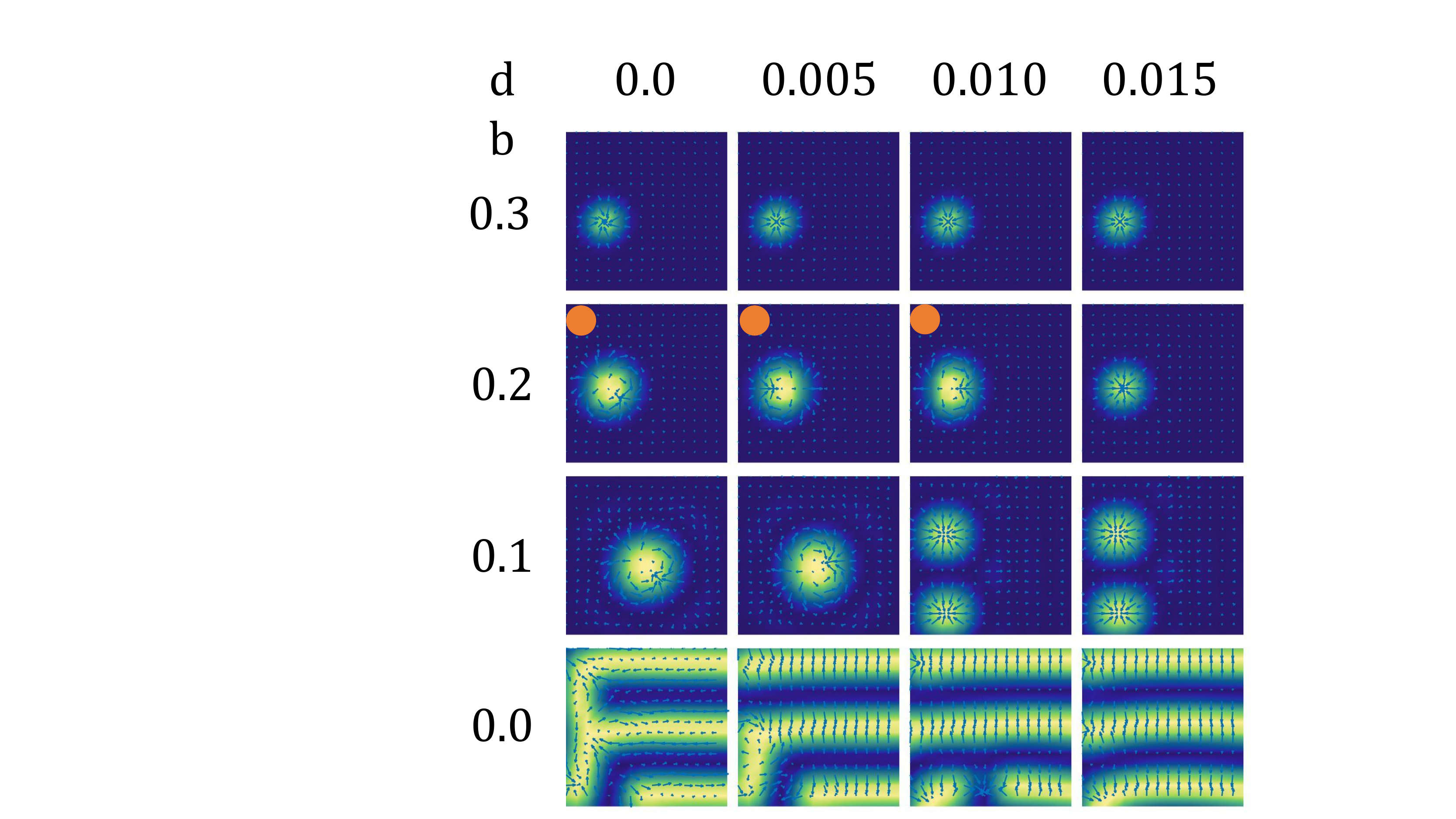}
		\caption{}
		\label{fig:bd_Q2}
	\end{subfigure}
	\begin{subfigure}[t]{.45\textwidth}	
			\adjincludegraphics[width=1\textwidth,trim={{.3\width} 0 0 {.05\height}},clip]{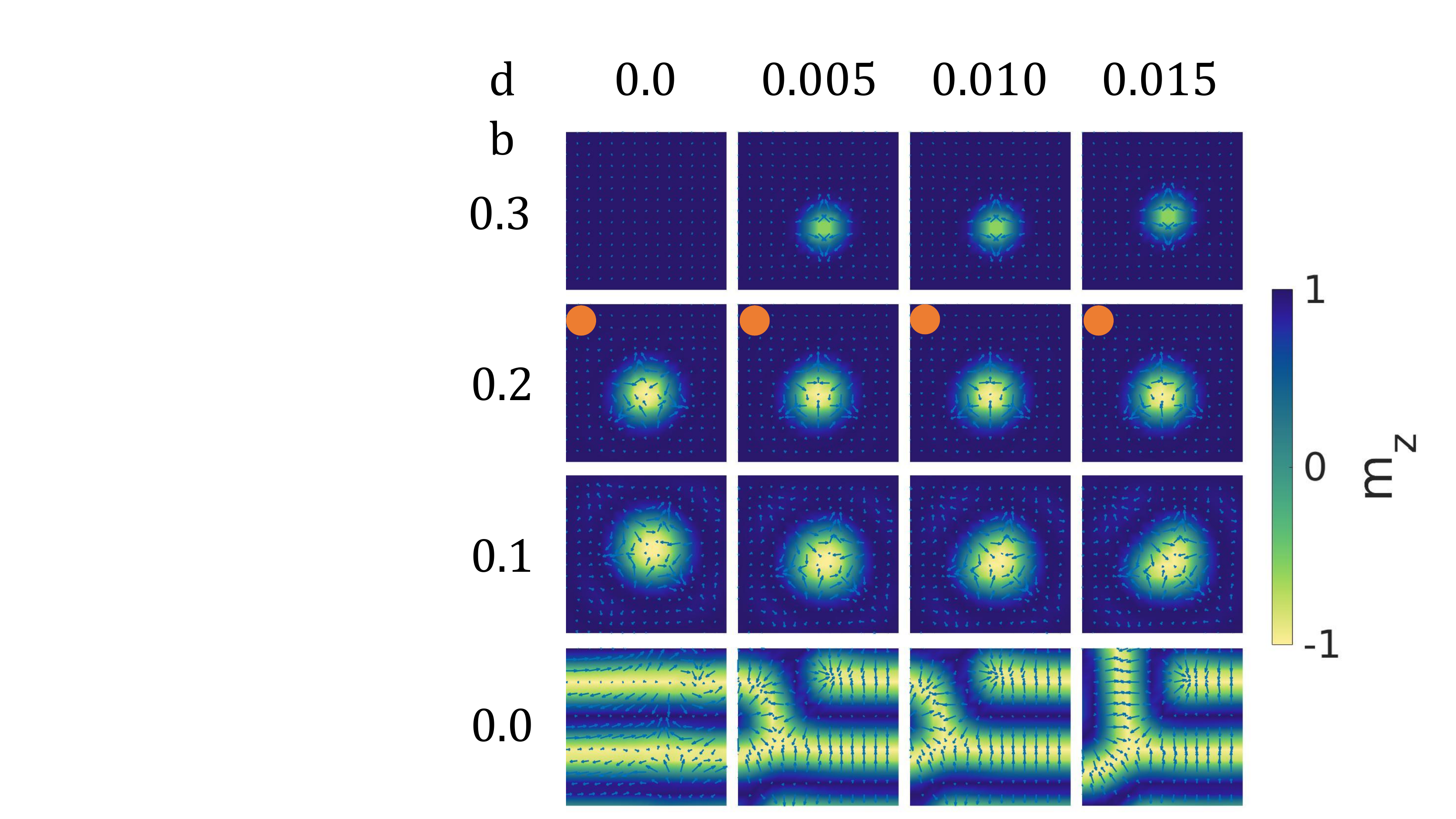}
		\caption{}
		\label{fig:bd_Qm2}
	\end{subfigure}
		
\caption{For $k=0.1$ and different values of the reduced DMI constant $d$ and the reduced applied field $b$, existence of (a) an isolated skyrmion solution, (b) an isolated antiskyrmion solution, (c) an isolated second-order skyrmion solution, (d) an isolated second-order antiskyrmion solution. Only metastable solutions of interest corresponding to excitations of the FM ground state are marked with an orange dot. The spin maps are zoomed in to show the topological defects. 
}
\label{fig:bd_diagram}
\end{figure}
  
 
\section{Results}\label{sec:results}
\subsection{Coexistence of skyrmion and antiskyrmion solutions}\label{subsec:coexistence}
We introduce a weak DMI to the system in order to break invariance with respect to the rotation of helicity, as well as the degeneracy between skyrmions and antiskyrmions. We look for a set of parameters that allow metastable skyrmion and antiskyrmion solutions to co-exist. To do so, we initialize the system close to a(n) (anti)skyrmion state and run an iterative energy minimization procedure similar to the one described in \cite{aharoni2000introductionminalgo}. If a(n) (anti)skyrmion solution exists, it is relaxed. We vary the perpendicular applied field $b$ in $[0, 0.4]$, and the DMI coupling constant $d$ in $[0, 0.07]$. We obtain the diagrams of Fig. \ref{fig:bd_diagram} for skyrmions [Fig. \ref{fig:bd_sk}] and antiskyrmions [Fig. \ref{fig:bd_antisk}]. 

At low field or high DMI, a spiral state is relaxed. Above a critical applied field, the spin-spiral state becomes the fully polarized  FM state via a second-order phase transition \cite{lin2016ginzburg}. From that state, metastable skyrmion solutions can be stabilized. In the presence of a perpendicular easy-axis anisotropy, the skyrmion lattice (SkX) becomes thermodynamically stable at intermediate applied field  \cite{lin2016ginzburg}. In this region, isolated skyrmions are more favorable than the FM state. They are metastable in the sense that they are local energy solutions, but not the ground state. At finite temperature, the isolated skyrmion state is rapidly destroyed by the nucleation of many topological defects in its vicinity. By contrast, the metastable solutions we are interested in are isolated skyrmionic defects as excitations of the FM ground state, which we mark as orange dots on Fig. \ref{fig:bd_diagram}. Since the symmetry of the DMI favors Néel skyrmions, degeneracies are lifted, and antiskyrmions become progressively less stable than their skyrmion counterparts as the DMI increases. They also relax diagonally on the square lattice, such that their Bloch-type axes are along the first-neighbor axes, which are the only ones that are DMI-coupled. This is due to the fact that one of the Néel-type axes of the antiskyrmion exhibits the opposite chirality to the one favored by the DMI. This introduces frustration in the orientation of the antiskyrmion solutions, which is not present in first-order, radially-symmetric skyrmion solutions.

Finally, Figs. \ref{fig:bd_Q2} and  \ref{fig:bd_Qm2} show the range of existence of second-order skyrmion and antiskyrmion solutions, respectively. They exist as stable solutions at low $d$ in the skyrmion lattice phase. By increasing the DMI at constant applied field, a bound skyrmion pair is relaxed instead of the second-order skyrmion, but the second-order antiskyrmion solution persists. Metastable solutions are found within a small window of the FM phase at sufficiently low field and DMI. We find second-order skyrmion and antiskyrmions to be quasi-degenerate in energy whenever they both exist, with small differences most likely caused by the underlying lattice. However, the range of existence of the antiskyrmion appears to be wider, as the first-order antiskyrmions are unfavored by the DMI.


\subsection{Paths to annihilation}\label{subsec:paths}




\begin{figure}[hbtp]
\centering	
	\begin{subfigure}[t]{.99\textwidth}	
		\adjincludegraphics[width=1\linewidth,trim={0cm 0cm 10cm 0cm},clip]{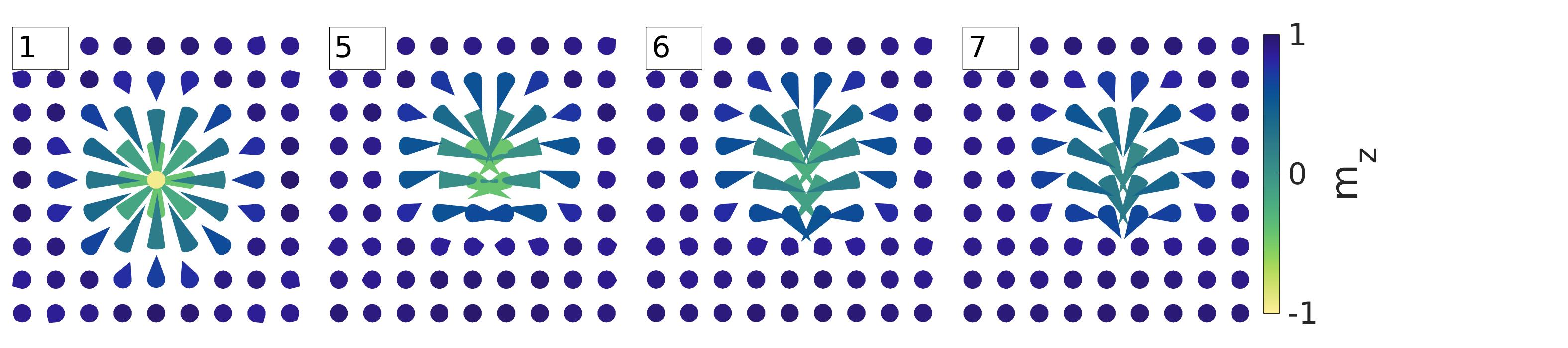}		
		\adjincludegraphics[width=1\linewidth,trim={0cm 0cm 10cm 0cm},clip]{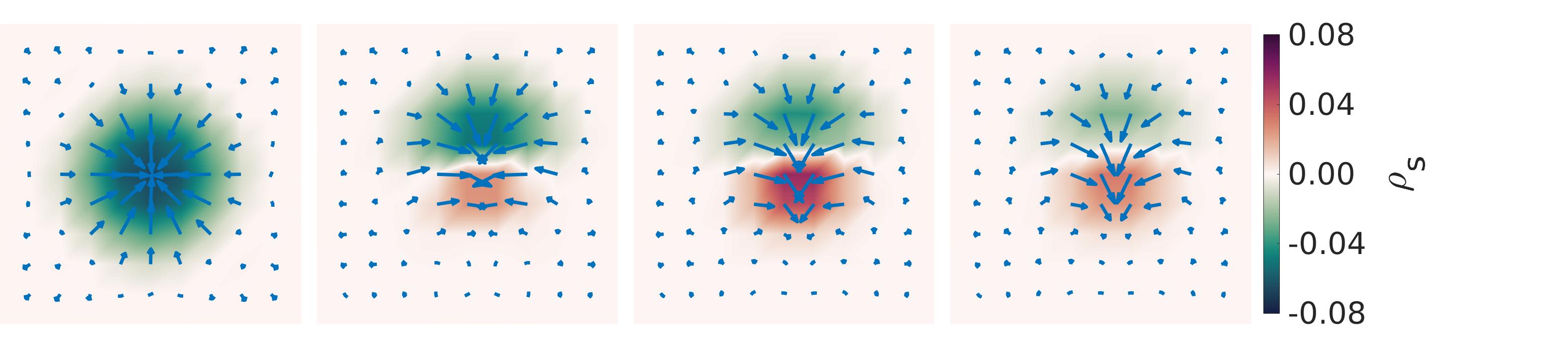}
		\caption{}
		\label{fig:sk_anni_d0d03}
	\end{subfigure}

		\begin{subfigure}[t]{.99\textwidth}	
		\adjincludegraphics[width=1\linewidth,trim={0cm 0cm 10cm 0cm},clip]{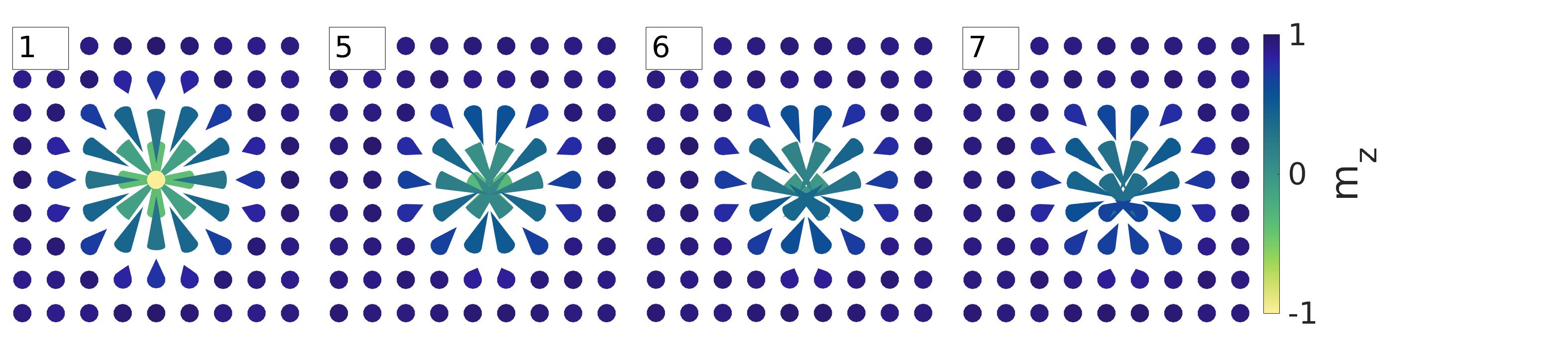}		
		\adjincludegraphics[width=1\linewidth,trim={0cm 0cm 10cm 0cm},clip]{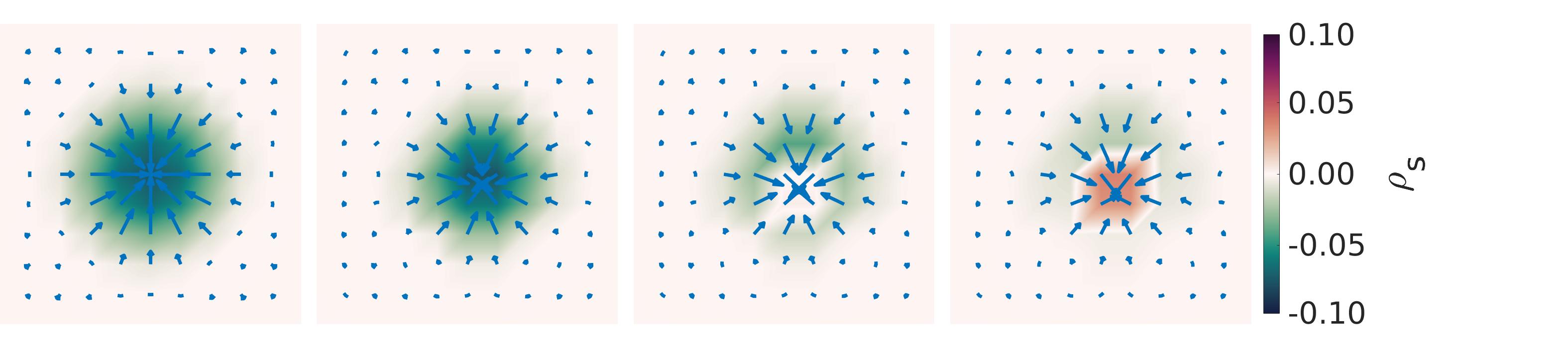}
		\caption{}
		\label{fig:sk_anni_d0d07}
	\end{subfigure}
	
		\begin{subfigure}[t]{.99\textwidth}	
	\adjincludegraphics[width=1\linewidth,trim={0cm 0cm 10cm 0cm},clip]{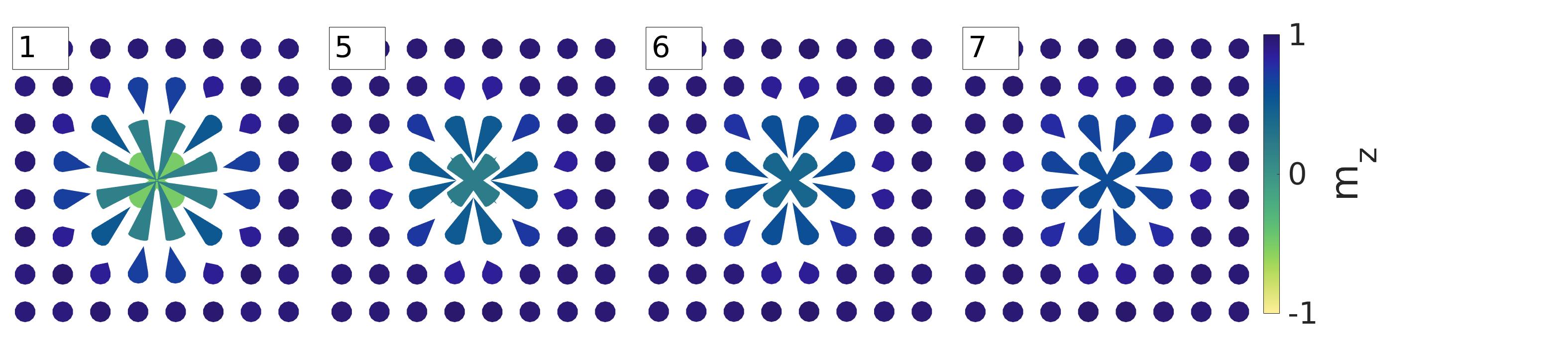}		
		\adjincludegraphics[width=1\linewidth,trim={0cm 0cm 10cm 0cm},clip]{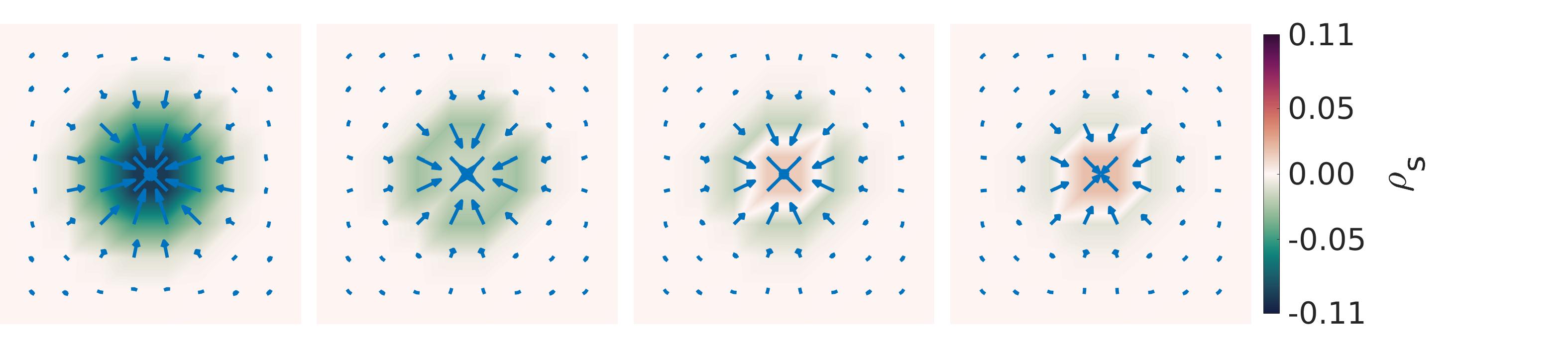}\hfill
		\caption{}
		\label{fig:sk_anni_d0d2}
	\end{subfigure}	
\caption{Spin maps (zoomed) and corresponding topological charge density along the transition path for a skyrmion annihilation. The parameters are  (a) $(b,d) = (0.2, 0.03)$ (metastable antiskyrmion solutions exist), (b) $(b,d) = (0.3, 0.07)$ (close to the existence of antiskyrmion solutions),  (c) $(b,d) = (0.7, 0.2)$ (antiskyrmions solutions do not exist). The image index is given in the top left-hand corner.
}
\label{fig:sk_paths}
	\end{figure}

	\begin{figure}[hbtp]
	\centering	
		\includegraphics[width=.7\textwidth]{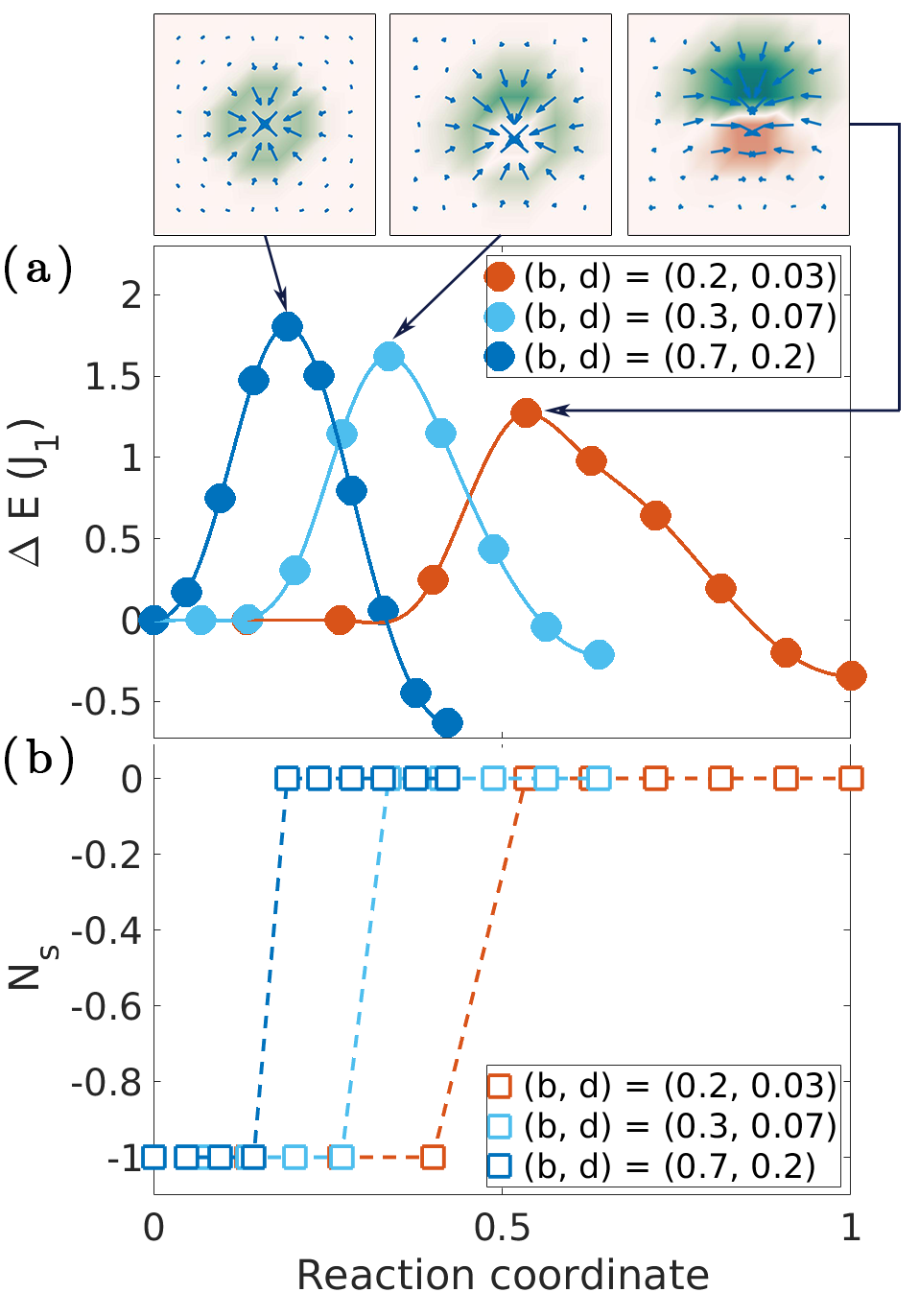}
	\caption{For the skyrmion annihilation at different values of the reduced applied field $b$ and reduced DMI constant $d$, we show the evolution along the transition path of (a) the internal energy barrier in units of $J_1$, (b) the topological charge. The reaction coordinate is normalized by the largest path length. The insets show the spin configuration and the topological charge density at the SP.\label{fig:etot_sk}}
\end{figure}

	
	\begin{figure}[hbtp]
\centering	
			\begin{subfigure}[t]{.99\textwidth}		
		\adjincludegraphics[width=1\linewidth,trim={0cm 0cm 10cm 0cm},clip]{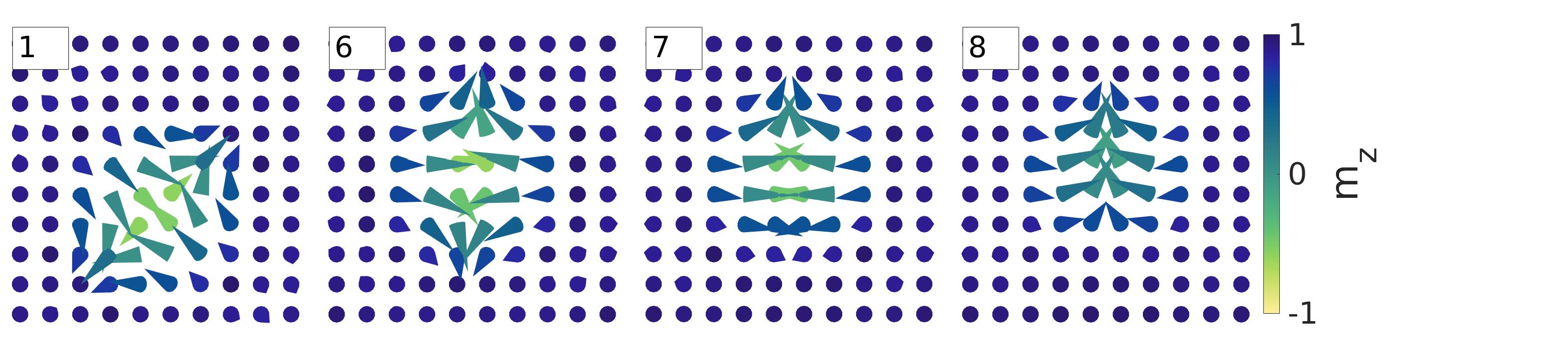}			
		\adjincludegraphics[width=1\linewidth,trim={0cm 0cm 10cm 0cm},clip]{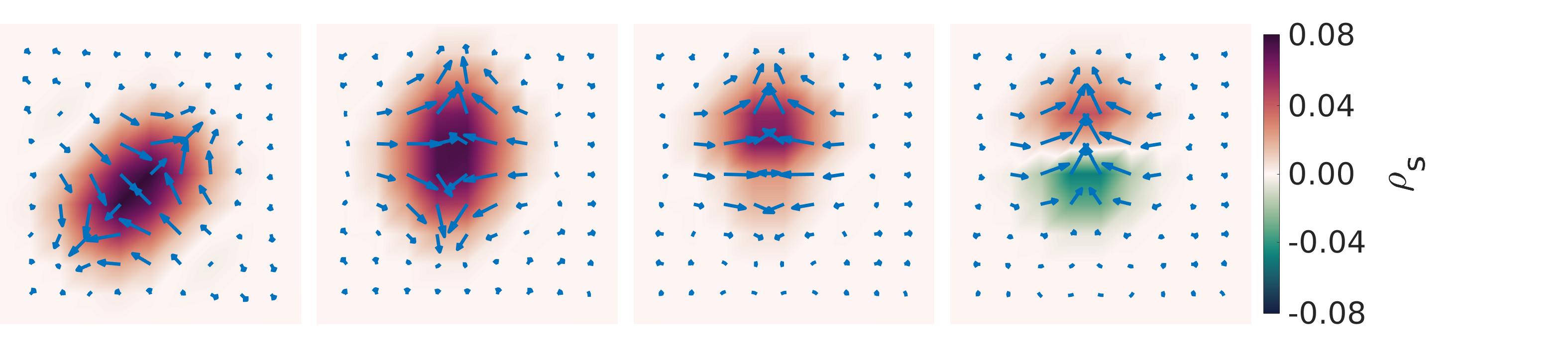}
		\caption{}
		\label{fig:antisk_anni_d0d03_b0d2_SP2}
	\end{subfigure}
	
	\begin{subfigure}[t]{.99\textwidth}		
		\adjincludegraphics[width=1\linewidth,trim={0cm 0cm 10cm 0cm},clip]{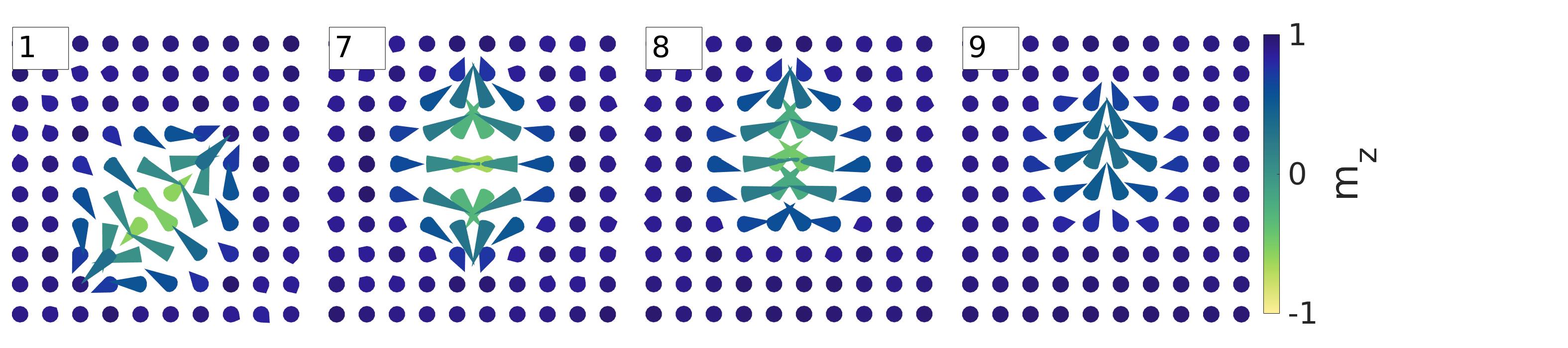}			
		\adjincludegraphics[width=1\linewidth,trim={0cm 0cm 10cm 0cm},clip]{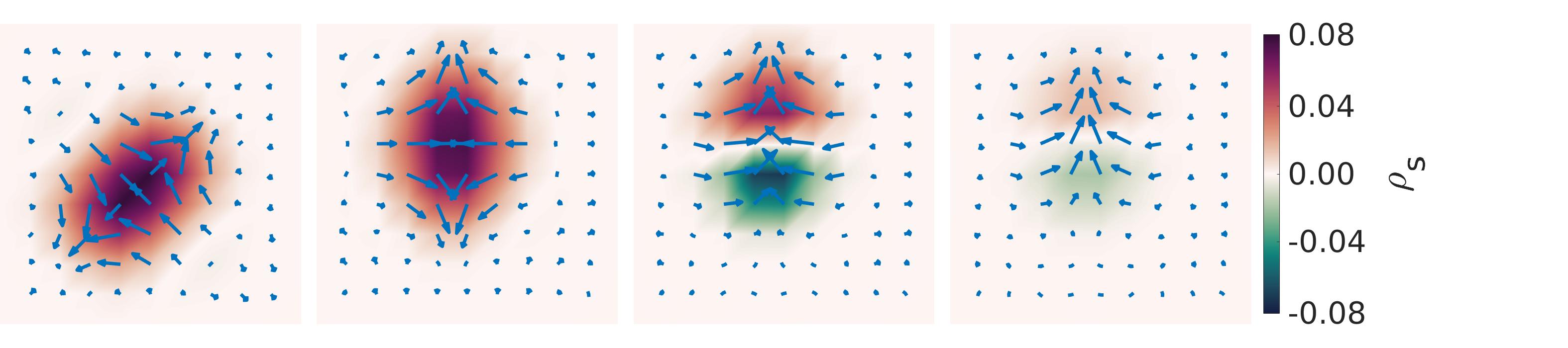}
		\caption{}
		\label{fig:antisk_anni_d0d03_b0d2_SP1}
	\end{subfigure}
	\caption{Spin maps (zoomed) and corresponding topological charge density along the transition path for an antiskyrmion annihilation with  $(b,d) = (0.2, 0.03)$ where (a) shows the path over SP$_1$ and (b) shows the path over SP$_2$. The image index is given in the top left-hand corner. \label{fig:antisk_paths}}
	\end{figure}
	

	\begin{figure}[hbtp]
\centering
\begin{subfigure}[t]{.7\textwidth}		
		\includegraphics[width=1\textwidth]{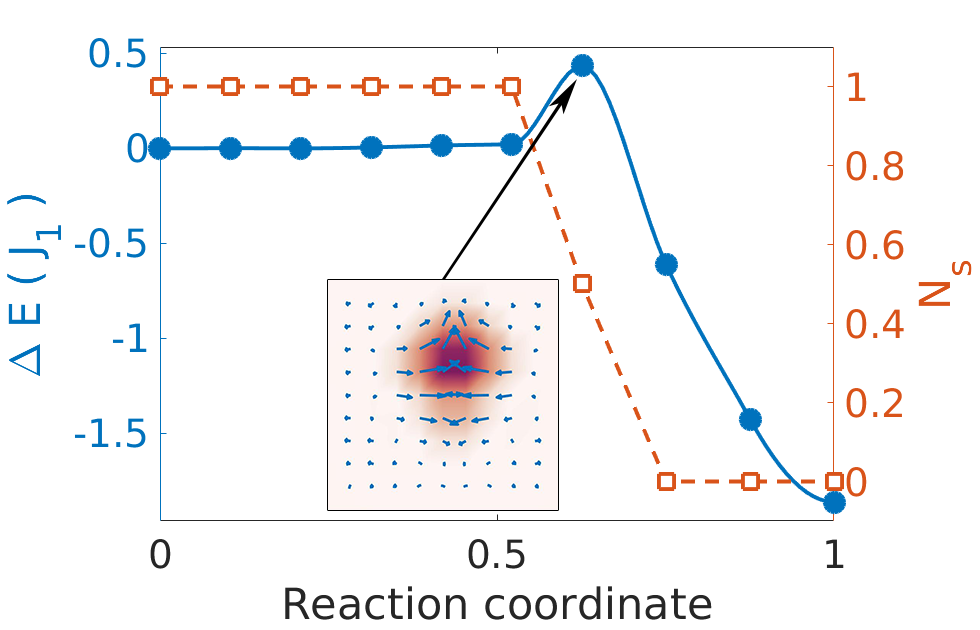}
		\caption{}
		\label{fig:etot_antisk_anni_SP2}
	\end{subfigure}
	
	\begin{subfigure}[t]{.7\textwidth}		
	\includegraphics[width=1\textwidth]{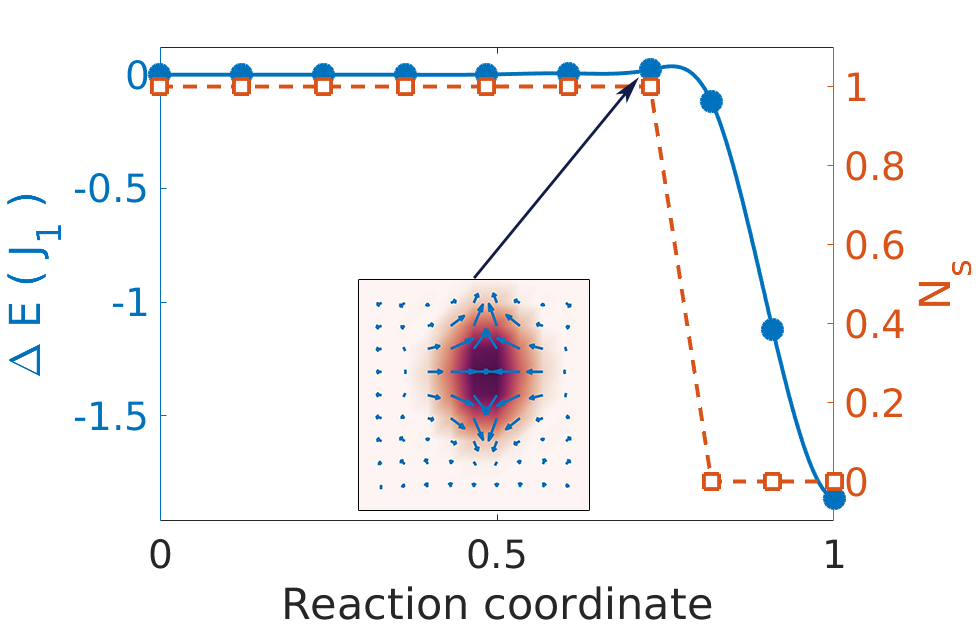}
		\caption{}
		\label{fig:etot_antisk_anni_SP1}
	\end{subfigure}
		\caption{Internal energy barrier in units of $J_1$ (filled blue dots) and topological charge $N_s$ (unfilled red squares) along the transition path for an antiskyrmion collapse with $(b,d) = (0.2, 0.03)$ for (a) the path over SP$_1$, (b) the path over SP$_2$. The inset shows the spin configuration and the topological charge density at the SP.}
	\label{fig:etot_antisk}

		\end{figure}

		\begin{figure}[hbtp]

	\begin{subfigure}[t]{.99\textwidth}		
		\adjincludegraphics[width=1\linewidth,trim={0cm 0cm 10cm 0cm},clip]{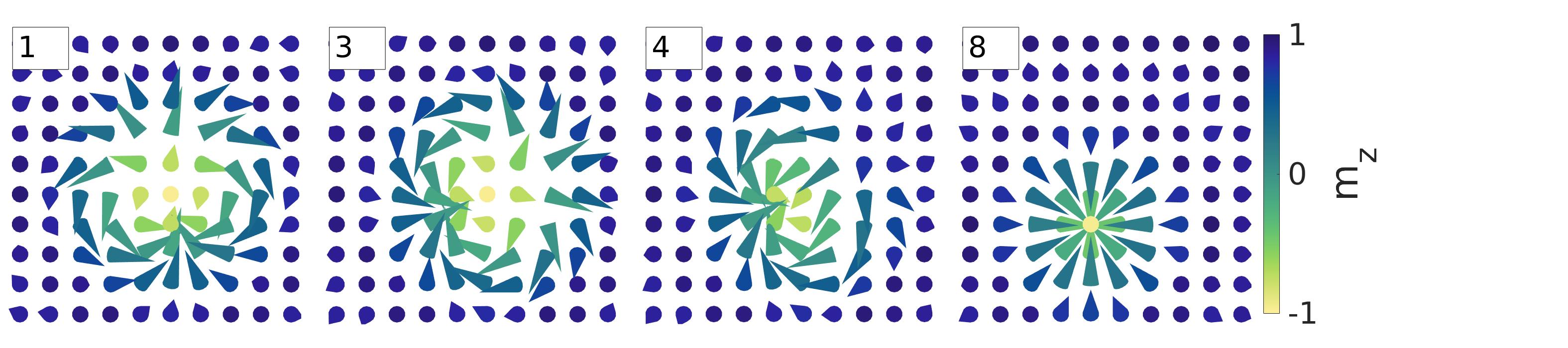}				
		\adjincludegraphics[width=1\linewidth,trim={0cm 0cm 10cm 0cm},clip]{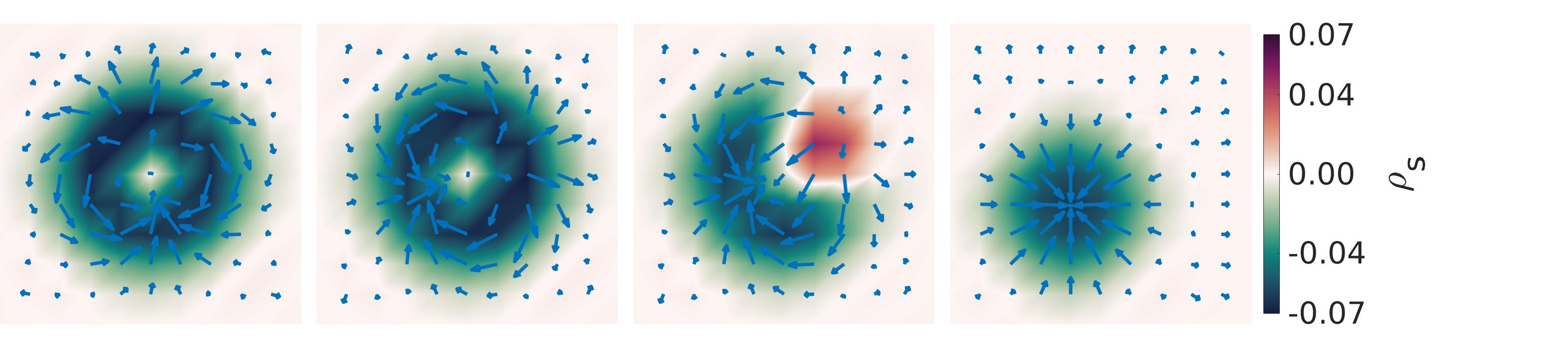}
						\caption{}
		\label{fig:Q2_Q1}
	\end{subfigure}
	
		\begin{subfigure}[t]{.99\textwidth}		
		\adjincludegraphics[width=1\linewidth,trim={0cm 0cm 10cm 0cm},clip]{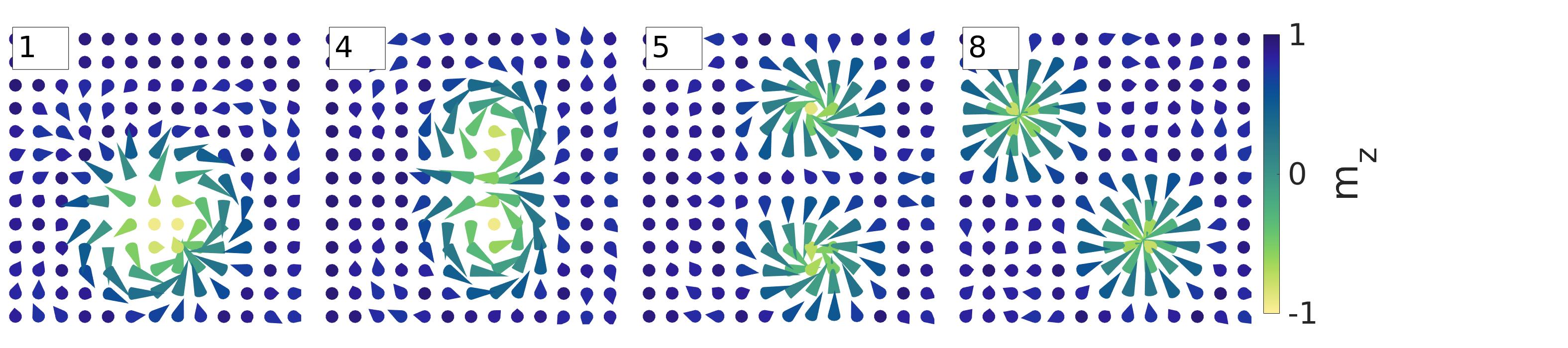}			
		\adjincludegraphics[width=1\linewidth,trim={0cm 0cm 10cm 0cm},clip]
		{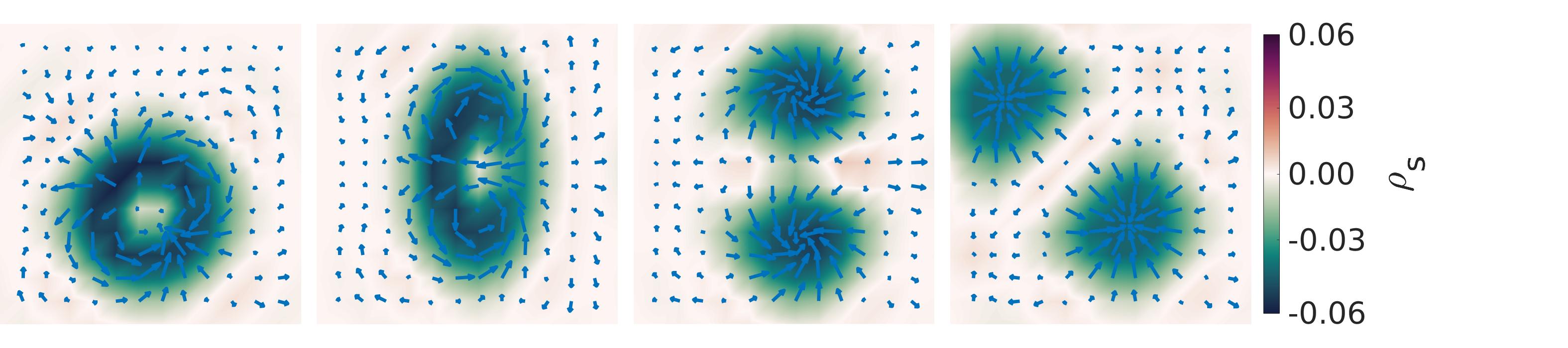}
						\caption{}
		\label{fig:Q2_Q1_Q1}
	\end{subfigure}
	
	\caption{Spin maps (zoomed) and corresponding topological charge density along the transition path for (a) the decay of a second-order skyrmion into a first-order skyrmion with $(b,d) = (0.14, 0.005)$,  (b) the division of a second-order skyrmion into a bound skyrmion pair with $(b,d) = (0.1, 0.005)$. The image index is given in the top left-hand corner. \label{fig:Q2_paths}}
	
		\end{figure}

		\begin{figure}[hbtp]
\centering
\begin{subfigure}[t]{.7\textwidth}		
		\includegraphics[width=1\textwidth]{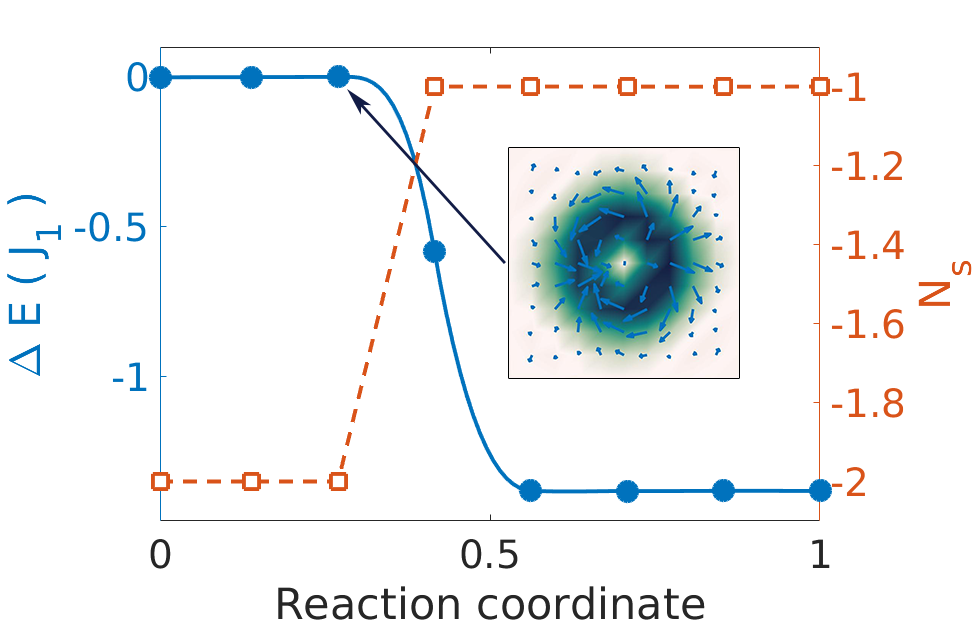}
		\caption{}
		\label{fig:etot_Q2_Q1}
	\end{subfigure}

\begin{subfigure}[t]{.7\textwidth}		
		\includegraphics[width=1\textwidth]{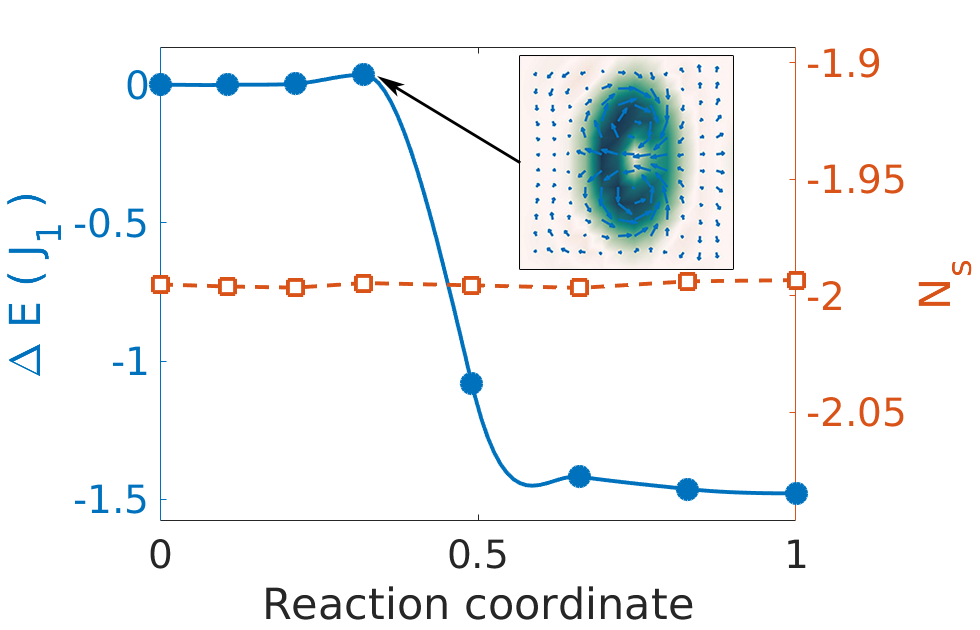}
		\caption{}
		\label{fig:etot_Q2_Q1_Q1}
	\end{subfigure}
			\caption{Internal energy barrier in units of $J_1$ (filled blue dots) and topological charge $N_s$ (unfilled red squares) along the transition path for (a) the decay of a second-order skyrmion into a first-order skyrmion with  $(b,d) = (0.14, 0.005)$, (b)  the division of a second-order skyrmion into a bound skyrmion pair with  $(b,d) = (0.1, 0.005)$. The inset shows the spin configuration and the topological charge density at the SP.}
	\label{fig:etot_Q2}
	\end{figure}

	
Transition mechanisms correspond to minimum energy paths (MEPs) on the $2N$-dimensional energy surface. We compute them via geodesics nudged elastic bands (GNEB) \cite{gneb} calculations with a climbing image (CI-GNEB) \cite{climbingimage} scheme to precisely identify the first-order saddle point (SP) on the path. Successive states of the system along the reaction coordinate are referred to as images. For each mechanism, we plot the spin maps at selected images and the corresponding topological charge density as defined in Eq. (\ref{eq:topo_charge_dens}) [Figs. \ref{fig:sk_paths}, \ref{fig:antisk_paths}, \ref{fig:Q2_paths}]. The total energy along the path is plotted in Figs. \ref{fig:etot_sk}, \ref{fig:etot_antisk}, and \ref{fig:etot_Q2} in units of the isotropic exchange coupling constant between first neighbors, $J_1$, as well as the total topological charge at each image, as defined in Eq. (\ref{eq:topo_charge}). 

\paragraph{Skyrmion.}The first set of parameters we examine is $(b,d) = (0.2, 0.03)$. It is a region of parameter space where both metastable skyrmions and antiskyrmions solutions exist. In this context, we observe an annihilation mechanism which is different from the usually reported isotropic type of collapse [Fig. \ref{fig:sk_anni_d0d03}]. One half of the skyrmion unfolds, and a half-antiskyrmion, or antimeron, is nucleated in its place. The injection of the opposite topological charge in the system constitutes the first-order saddle point on the transition path [Fig. \ref{fig:etot_sk}]. This state possesses four different realizations, equivalent to rigid $\pi/2$ rotations of the whole sample. The remaining meron and antimeron then naturally annihilate.  We found that the isotropic type of collapse also exists in this system, but it involves a higher order SP and therefore constitutes a less probable route.  
The second set of parameters is  $(b,d) = (0.3, 0.07)$. In the space of parameters, it is situated just above the limit at which antiskyrmion solutions no longer exist. In this case, the skyrmion undergoes a regular collapse [Fig. \ref{fig:sk_anni_d0d07}] involving a first-order SP (image 6), but the core displays a weak asymmetry in a way that is reminiscent of Fig. \ref{fig:sk_anni_d0d03}. Since we are close to the region of antiskyrmions' metastability, this mechanism can be considered as intermediate between antimeron nucleation and isotropic collapse.
Finally, we select a region of parameter space with high DMI, far from the metastability region of antiskrymions:  $(b,d) = (0.7, 0.2)$. As expected, we obtain a perfectly isotropic collapse [Fig. \ref{fig:sk_anni_d0d2}] without the injection of an opposite topological charge. In all cases, the SP configuration corresponds to an almost zero topological charge, and the variation of the topological charge along the path is more abrupt with increasing DMI.

 \paragraph{Antiskyrmion.} We revert back to the previous set of parameters, $(b,d) = (0.2, 0.03)$, which allows metastable antiskyrmions to coexist with skyrmions. We look at the path to annihilation of an antiskyrmion and we relax two different mechanisms passing through a first-order SP [Figs. \ref{fig:antisk_anni_d0d03_b0d2_SP1} and \ref{fig:antisk_anni_d0d03_b0d2_SP2}]. In both cases, the annihilation process begins with the rotation of the antiskyrmion, such that its Néel-type axes are aligned along the first-neighbor axes (images 1-7).  This configuration constitutes an energy maximum as it is unfavored by the DMI, and the antiskyrmion can reach this state either as a full antiskyrmion with full integer topological charge (SP$_2$ on Figs. \ref{fig:antisk_anni_d0d03_b0d2_SP2} and \ref{fig:etot_antisk_anni_SP2}), or the spins may begin to unwind, which leads to a decrease in the topological charge (SP$_1$ on  Figs. \ref{fig:antisk_anni_d0d03_b0d2_SP1} and \ref{fig:etot_antisk_anni_SP1}). Past the SP, similarly to the annihilation mechanism of its skyrmion counterpart,  the topological charge drops as a meron is nucleated and annihilates with the remaining antimeron. SP$_1$ exists in four possible realizations ($\pi/2$-rotations), while SP$_2$ exists in two ($\pi$-rotation). It is interesting to note that the fact that the antiskyrmion is frustrated in its orientation on the lattice suffices to distinguish between a metastable solution and a SP. Here, the coupling to the spin lattice entails that the antiskyrmion, as well as the second-order skyrmion and all non-radially symmetric solutions behave as strongly correlated spin-ice \cite{yavors2008dy}.

 \paragraph{Second-order skyrmion.} We select paramaters that allow both the first- and second-order skyrmions to be metastable: $(b,d) = (0.14, 0.005)$. 
We initially set the final state of the GNEB calculation to a Néel skyrmion, and we obtain the mechanism in Fig. \ref{fig:Q2_Q1}, with the corresponding energy profile shown on Fig. \ref{fig:etot_Q2_Q1}. The second-order skyrmion first rotates on the lattice and reaches the SP (image 3). The nucleation of an antimeron along the Néel axis follows (image 4) and the energy drops, accompanied by an abrupt change in the topological charge. The antimeron subsequently annihilates with part of the second-order skyrmion, leaving a first-order Néel skyrmion in its place.
Another route to annihilation for the second-order skyrmion
corresponds to its division into a bound skyrmion pair, as shown on Fig. \ref{fig:Q2_Q1_Q1}. The bound skyrmion pair is a metastable solution, since the interaction potential of skyrmions in frustrated magnets is non-monotonic as a function of distance, and is found in turn positive (repulsive) and negative (binding) \cite{leonov2015multiply,lin2016ginzburg}. 
This mechanism does not involve any significant change in the topological charge of the total system [Fig. \ref{fig:etot_Q2_Q1_Q1}], but instead a redistribution of the topological charge density. Both of the above mechanisms possess four realizations, corresponding to $\pi/4$ rigid rotations of the sample. As for the fate of the skyrmion pair, the binding potential is quite shallow compared to the activation barrier for annihilation (from the results of \cite{lin2016ginzburg}, we can estimate the unbinding barrier to be $\sim 10^{-4}$ $J_1$). Therefore, the skyrmions will in most cases separate before annihilating individually.

Similar processes where reported in \cite{zhang2017skyrmion} from dynamics simulations, in which the second-order skyrmion division was triggered by both a driving current at zero-temperature, and thermal fluctutations at finite temperature. The finite temperature simulations also showed the decay of the second-order skyrmion into a first-order skyrmion.


\subsection{Langevin simulations}

\begin{figure}
\centering
	\includegraphics[width=1\textwidth]{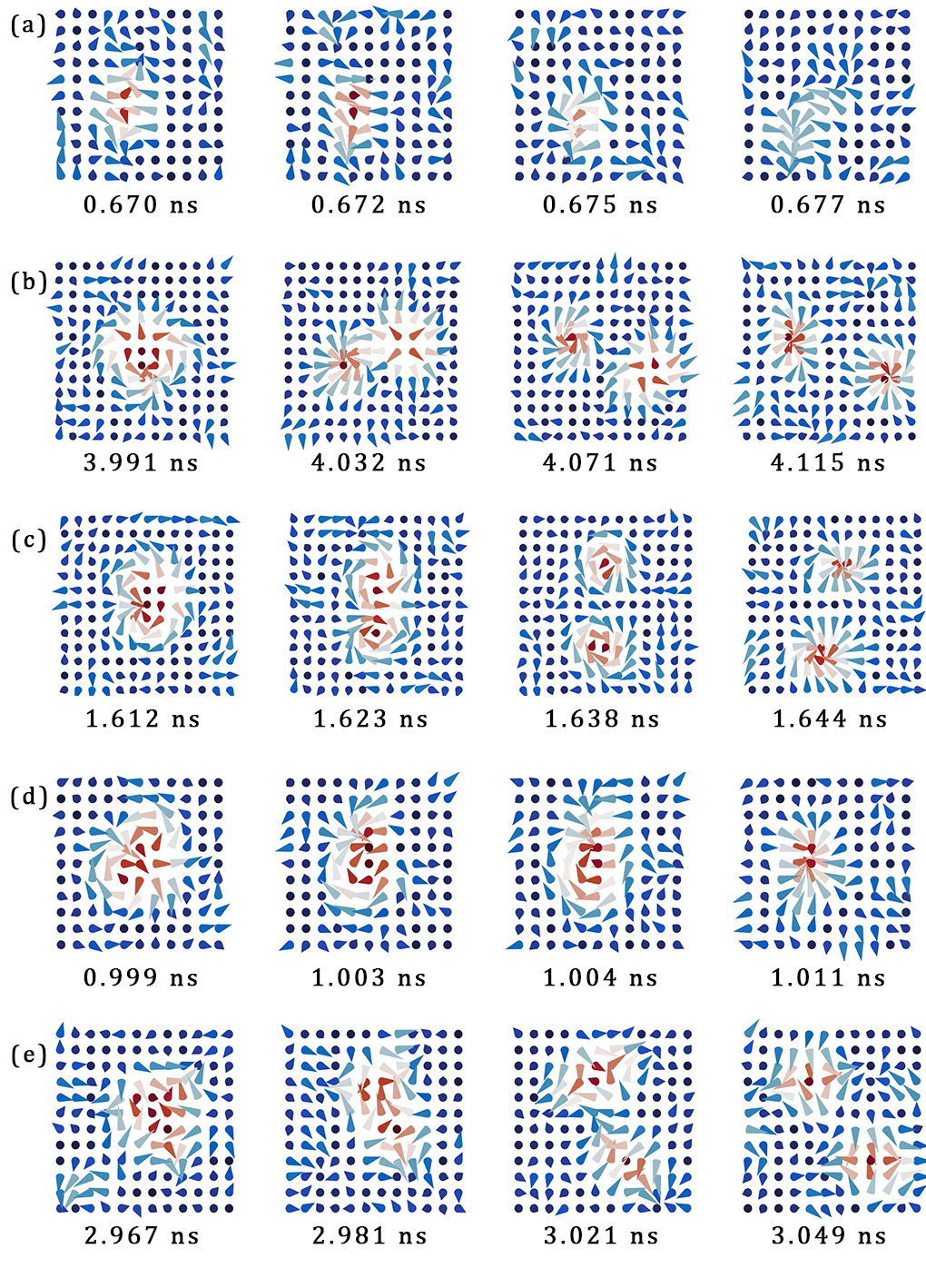}
\includegraphics[width=.2\textwidth]{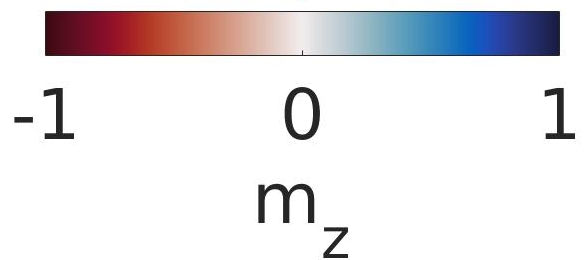}
\caption{Langevin simulation snapshots (zoomed) of (a) the annihilation of an antiskyrmion at $T=$ 80 K, $(b,d) = (0.2, 0.03)$ and,
at $T = 50$ K, ($b, d$) = (0.14, 0.005), (b) and (c) the division of a
second-order skyrmion into a bound skyrmion pair, (d) the decay
of a second-order skyrmion into a first-order skyrmion, and (e) the
division of a second-order antiskyrmion into a bound antiskyrmion
pair.}
\label{fig:langevin}
\end{figure}

In order to confirm the previously calculated MEPs, we perform direct atomistic Langevin simulations at low temperature. Details on the method can be found in Appendix \ref{app:langevin}. We observe annihilation mechanisms for which we show snapshots of the spin configuration in Fig. \ref{fig:langevin}.
 For the antiskyrmion at $T=$ 80 K and $(b,d) = (0.2, 0.03)$ [Fig. \ref{fig:langevin}a], we observe the meron-nucleation type of annihilation that we reported on Fig. \ref{fig:antisk_paths}. Since we have short lifetimes within the nanosecond (ns) timescale, we can estimate the antiskyrmion's stability directly from Langevin simulations. We use the topological charge of the system to track the collapse of the antiskyrmion. We record 390 collapses and obtain an average lifetime $\tau_{\text{DL}}(80$K$) = 0.49$ ns, with a standard deviation of the order of the lifetime, $\sigma = 0.46$ ns.
  
 For the second-order skyrmion and antiskyrmion, as they overall exhibit lower activation energies, we perform simulations at $T=50$ K, and $(b,d) = (0.14, 0.005)$. The division of the second-order skyrmion into a bound skyrmion pair seems to occur along at least two different paths shown on Fig. \ref{fig:langevin}b and \ref{fig:langevin}c, in which the latter corresponds to the MEP on Fig. \ref{fig:Q2_Q1_Q1}. In Fig. \ref{fig:langevin}b, the division occurs through excitation of one half of the second-order skyrmion, which eventually leads to the separation of the two halves. On the other hand, in Fig. \ref{fig:langevin}c, we witness a more symmetric kind of division.
For the same set of parameters, we also observe the decay into a single skyrmion [Fig. \ref{fig:langevin}d, which corresponds to the path on Fig. \ref{fig:Q2_Q1}]. Finally, on Fig. \ref{fig:langevin}e, we show the division of a second-order antiskyrmion into a pair of first-order antiskyrmions.


\subsection{Annihilation rates}\label{subsec:stability}
The rate of thermally activated processes with an internal energy barrier $\Delta E$ can be described by the Arrhenius law, $ f(T) = f_0 e^{-\Delta E /k_B T}.$
We use a form of Langer's theory for the decay of metastable states \cite{langer} adapted to magnetic spin systems \cite{coffey, seuss, desplat2018thermal} to compute individual rate prefactors $f_0$ for each of the mechanisms described previously, while the CI-GNEB scheme directly gives us a precise value for the activation energy. More details on the methods in this paragraph are given in Ref. \onlinecite{desplat2018thermal}. The results are gathered in Table \ref{tab:transition_rates}. 
To account for all equivalent realizations of a given mechanism (rigid rotations of the spin configurations at the SP), the prefactor $f_0$ is multiplied by a factor two or four when needed, in accordance with Sec. \ref{subsec:paths}. 

Based on these results, we can estimate the lifetime of the antiskyrmion calculated with Langer's theory (with the subscript TST, for transition state theory), and compare it with the average lifetime previously obtained from direct Langevin simulations. Assuming independent processes, we have $f_{\text{TST}}(80$K$) = f_{01}e^{-\beta \Delta E_1} + f_{02}e^{-\beta \Delta E_2} = \tau_{\text{TST}}^{-1}(80$K$)= 2.08$ GHz and $\tau_{\text{TST}}(80$K$)$ = 0.48 ns. Although we are not completely within the scope of Langer's theory ($\beta \Delta E_1 = 0.3$, as opposed to the recommended $\beta \Delta E \geq 5$ \cite{coffey}), this shows an excellent agreement with the result from direct Langevin, $\tau_{\text{DL}}(80$K$)$ = 0.49 ns.  

In the case of the second-order skyrmion's decay into a first-order skyrmion, we find a Goldstone mode at both the metastable and transition states, which do not seem to clearly correspond to a translational or rotational invariance, hence we cannot give a definite value for the attempt frequency in this case.

\begin{table}
\centering
\caption{Internal energy barrier $\Delta E$ and rate prefactor $f_0$ for the annihilation of a skyrmion, an antiskyrmion, and a second-order skyrmion with different values of the reduced applied field $b$ and reduced DMI $d$. \label{tab:transition_rates}}
 \begin{ruledtabular}
\begin{tabular}{cccccccc}
 mechanism & $b$ ($J_1$) & $d$ ($J_1$) & $\Delta E$ ($J_1$) & $ f_0$ (GHz)   \\
\hline
   & & skyrmion &   & & &  \\
antimeron nucl. & 0.2 & 0.03 & 1.27 &  109.9 $\times$ 4  \\
iso. collapse & 0.3 & 0.07 & 1.62 &  3639.5  \\
iso. collapse & 0.7 & 0.2 & 1.80 &  1247.5  \\ 
\hline
 & &  antiskyrmion &   & & & \\
meron nucl. SP$_1$ & 0.2 & 0.03 & 0.022 &  1.4 $\times$ 2\\
meron nucl. SP$_2$ & 0.2 & 0.03 & 0.43 &   10.6 $\times$ 4   \\
\hline
  & & 2nd. order sk. &  &  & &   \\
antimeron nucl. & 0.14 & 0.005 & 0.0011 &   - \\ 
sk. pair division  & 0.14 & 0.005 & 0.062 & - \\

\end{tabular}
\end{ruledtabular}
\end{table}

\section{Discussion and Conclusion}\label{sec:discussion}

\subsection{Topological transitions: taking a ball out of a net}

\begin{figure}
\centering
\begin{subfigure}[t]{1\textwidth}	
	\includegraphics[width=.9\textwidth]{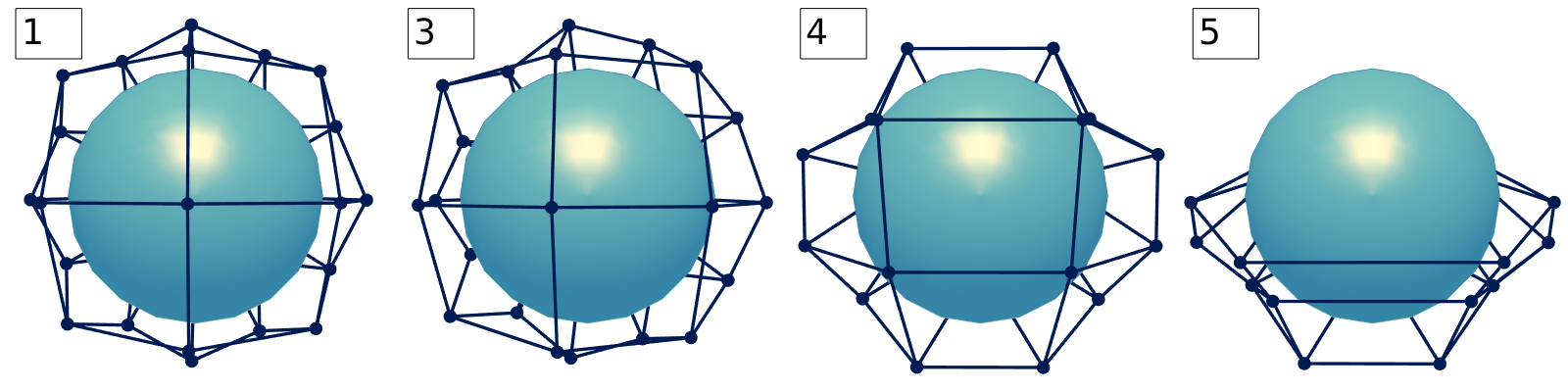}	
		\caption{}
		\label{fig:sk_anni_d0d03_spheres}
	\end{subfigure}
	
	\begin{subfigure}[t]{1\textwidth}	
		\includegraphics[width=.9\textwidth]{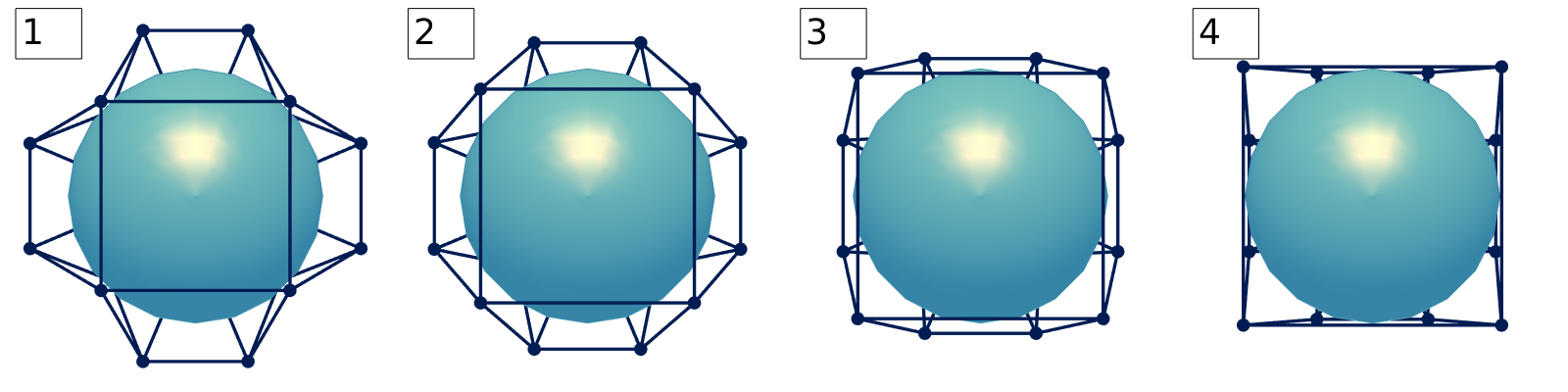}
		\caption{}
		\label{fig:sk_anni_d0d2_spheres}
	\end{subfigure}
	
		\begin{subfigure}[t]{1\textwidth}	
		\includegraphics[width=.9\textwidth]{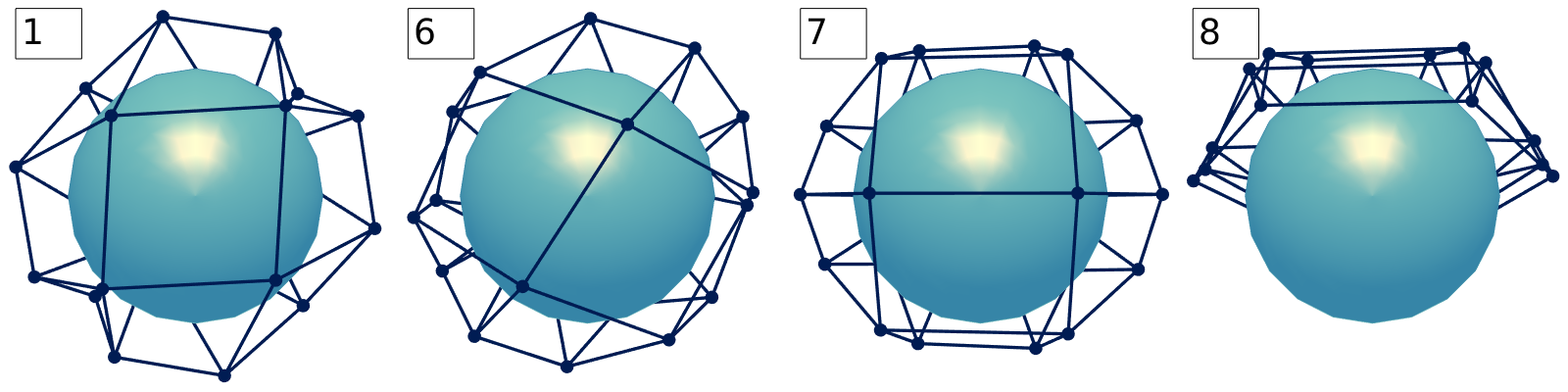}	
		\caption{}
		\label{fig:antisk_anni_SP1_spheres}
	\end{subfigure}
	
			\begin{subfigure}[t]{1\textwidth}	
		\includegraphics[width=.9\textwidth]{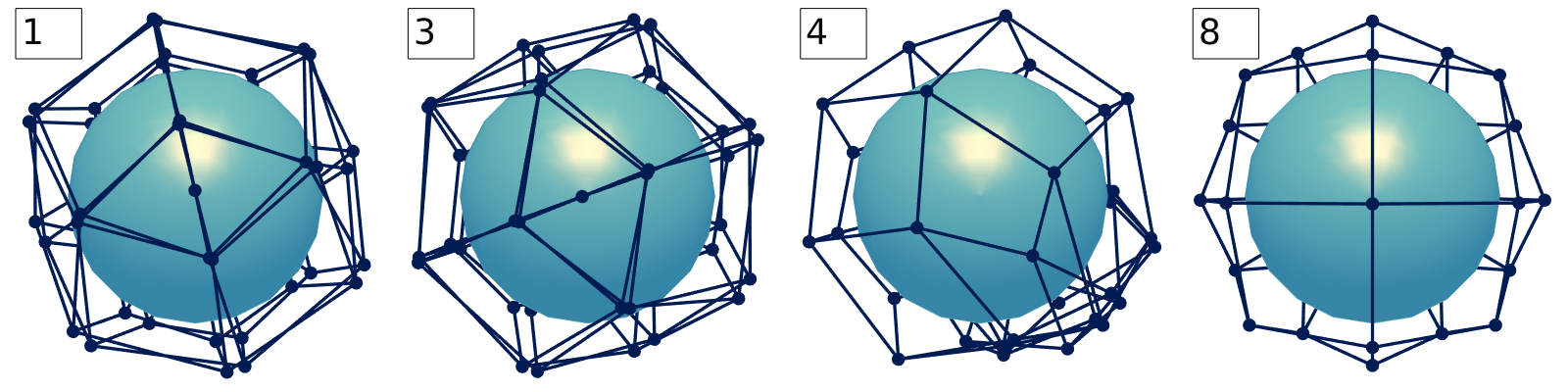}	
		\caption{}
		\label{fig:Q2_Q1_spheres}
	\end{subfigure}

	\caption{Mapping of the spin configurations along the MEP onto the unit sphere. Vertices correspond to the tip of the magnetic vectors with their origin in the center of the sphere, and the edges represent the exchange coupling between first neighbors.  The solid sphere represents the `ball' that gets extracted from the `net'. The image index is given in the top left-hand corner.  We show (a) a skyrmion annihilation via antimeron nucleation [Fig. \ref{fig:sk_anni_d0d03}], (b) a  skyrmion annihilation via isotropic collapse [Fig. \ref{fig:sk_anni_d0d2}], (c) an antiskyrmion annihilation via meron nucleation [Fig. \ref{fig:antisk_anni_d0d03_b0d2_SP2}]), (d) a second-order skyrmion decay into a first-order skyrmion [Fig. \ref{fig:Q2_Q1}].  \label{fig:spheres}}
	\end{figure}

Drawing inspiration from Ref. \onlinecite{rohart2016path}, we project images along the MEPs onto the space of configurations -- the unit sphere [Fig. \ref{fig:spheres}]. Vertices correspond to the tip of the magnetic vectors with their origin in the center of the sphere, and the edges represent the exchange coupling between first neighbors. The ferromagnetic background (spins along $+z$) is found at the north pole of the sphere, while the core of a skyrmion typically points along the south pole (along $-z$). The view is set just below the south pole, looking axially towards the $+z$ direction. Annihilating a topological defect to recover the uniformly magnetized state means bringing the mesh back to the north pole by moving the vertices on the sphere. This can be thought of as taking a ball (solid blue sphere) out of a net (dark blue mesh) by deforming the net \cite{rohart2016path}. 
Annhilations that occur via the injection of the opposite topological charge [Figs. \ref{fig:sk_anni_d0d03_spheres} and \ref{fig:antisk_anni_SP1_spheres}]  involve a rearranging of the spin-mesh in a way that allows the entire net to be removed along a single direction. By contrast, an isotropic collapse consists in progressively spreading the mesh open by an equal amount along all directions [Fig. \ref{fig:sk_anni_d0d2_spheres}]. Finally, a second-order skyrmion corresponds to a kind of double net [Fig. \ref{fig:Q2_Q1_spheres}]. Its decay into a first-order skyrmion via antimeron injection consists in the removal of one of the nets, again along a given direction, so that only a single net remains.

\subsection{Conclusion}
In regions of parameter space for which both metastable skyrmion and antiskyrmion solutions can be realized, the most probable paths to annihilation -- that is, the ones that go over a first-order saddle point -- seem to the paths involving the injection of the opposite topological charge into the system, in the form of the nucleation of merons and antimerons. The injection of the opposite charge usually happens just after the saddle point configuration, or at the saddle point in the case of the skyrmion. The reason these peculiar paths exist in the present system seem to be the interplay of the frustrated exchange, and the small sizes of the topological defects (small number of magnetic sites involved in the spin textures) that can be stabilized, which is directly linked with the period of the spin spiral ground state determined by the exchange frustration. Alternatively, the division of the second-order skyrmion into a bound skyrmion pair involves no change in the total toplogical charge of the system, only a redistribution of the charge density, and is also a valid path.


Overall, the skyrmionic solutions in this system are found to be rather unstable, as they exhibit low internal energy barriers, combined with many possible paths to collapse with attempt frequencies in the range of several GigaHertz (GHz). The existence of many possible paths is a direct consequence of the exchange frustration, which drastically complexifies the energy landscape and is responsible for the emergence of many (meta)stable and saddle point states, as well as many possible paths connecting them. The low stability allowed for a direct comparison of the average lifetime of the antiskyrmion at $80$ K computed from direct Langevin simulations, with transition state theory calculations, and we obtained a very good agreement. Nevertheless, in this context, the direct use of transition state theory to compile an overall reliable lifetime for any given structure seems ill-advised, as it is difficult to account for all possible mechanisms. As for the low activation energies and not particularly low attempt frequencies, we can once more relate them to the skyrmions being very small, and possessing only a few internal modes \cite{desplat2018thermal,von2019skyrmion}. 

\begin{acknowledgments}
This work was supported by the Horizon 2020 Framework
Programme of the European Commission, under Grant agreement
No. 665095 (MAGicSky), and the Agence Nationale de la Recherche under Contract No. ANR-17-CE24-0025 (TOPSKY). The authors thank B. Dupé for enlightening discussions, and D. Suess for previous help with code testing.  
\end{acknowledgments}

\appendix

\section{Simulation parameters}\label{app:parameters}
We use the following parameters for a two-dimensional monolayer of $40 \times 40$ simulated sites: the isotropic exchange constant between first neighbors is set to $J_{1} = 1.6\times 10^{-20}$ J ($\sim 100$ meV) with lattice constant $a=1$ nm and saturation magnetization $ M_s = 1.1 a^3 $ MA.m$^2$  \cite{thiaville2012dynamics}.
The gyromagnetic ratio is that of the free electron , $\gamma = 1.76 \times 10^{11}$ rad s$^{-1}$ T$^{-1}$, and the dimensionless damping factor is $\alpha = 0.5$  \cite{thiaville2012dynamics} (Intermediate-to-high damping regime of Langer's theory).

\section{Langevin dynamics} \label{app:langevin}
The dynamics of the magnetic spin system $\{\mathbf{m}_i\} $, $i=1\dots N$, is governed by the set of coupled, dimensionless, stochastic Landau-Lifshitz-Gilbert (LLG) equations \cite{garcia1998langevin}:

\begin{equation}\label{eq:stoch_dimless_LLG}
\frac{\diff \mathbf{m}_i}{\diff \bar{t}} = \frac{1}{\alpha} \mathbf{m}_i \times \big[ \mathbf{b}_\text{eff} + \mathbf{b}_\text{fl} (\bar{t}) \big] -
\mathbf{m}_i \times \big( \mathbf{m}_i \times \big[ \mathbf{b}_\text{eff} + \mathbf{b}_\text{fl} (\bar{t}) \big] \big).
\end{equation}
All the quantities are in units of the isotropic exchange coupling constant between first neighbors $J_1$, such that in the above expression, $\mathbf{b}_\text{eff} =  -\dfrac{1}{J_1} \dfrac{\partial E}{\partial \mathbf{m}_i}$ is the local reduced effective field, and $\mathbf{b}_\text{fl}$ is a stochastic fluctuating field in the form of white noise, which accounts for fluctuations of the orientation of $\mathbf{m}_i$ caused by interactions with microscopic degrees of freedom of the environment. 
It is assumed to be Gaussian distributed with the following statistical properties, in agreement with the fluctuation-dissipation theorem \cite{garcia1998langevin}: 
 \begin{equation}
 \begin{split}
 & \langle b_{\text{fl},j}(\bar{t})\rangle=0, \\
 &  \langle b_{\text{fl},j}(\bar{t})b_{\text{fl},k}(\bar{t}')\rangle=2 \Big( \frac{\alpha^2}{1+\alpha^2} \frac{k_B T}{J_1} \Big) \delta_{jk}\delta(\bar{t}-\bar{t}'),\\
 \end{split}
\end{equation}
where $j,k$ are Cartesian indices, $\langle \rangle$ denotes an average over many realizations of the fluctuating field, $\delta(\bar{t}-\bar{t}')$ is the Dirac distribution, and  $\delta_{jk}$ is the Kronecker symbol. The reduced time $\bar{t}$ is linked to physical time $t$ via $t = \dfrac{M_s }{ \alpha \gamma J_1} \bar{t}$, and we maintain $d\bar{t} = 0.001$. We use the stochastic Heun scheme \cite{garcia1998langevin}, which converges to the solution of (\ref{eq:stoch_dimless_LLG}) when the multiplicative noise is interpreted in the sense of Stratonovich. Additionally, periodic boundary conditions are applied to prevent the skyrmions from escaping through the boundaries.

\newpage
\bibliography{/home/louise/Dropbox/writing/latex/skyrmionbib}

\end{document}